\documentclass[aps,pra,twocolumn,showpacs,unsortedaddress]{revtex4-1}

\usepackage{amsmath,amssymb}
\usepackage{graphicx}
\usepackage{bbm}

\newcommand{\ii}{\mathrm i}

\usepackage{color}

\usepackage{booktabs}
\AtBeginDocument{
\heavyrulewidth=.08em
\lightrulewidth=.05em
\cmidrulewidth=.03em
\belowrulesep=.65ex
\belowbottomsep=0pt
\aboverulesep=.4ex
\abovetopsep=0pt
\cmidrulesep=\doublerulesep
\cmidrulekern=.5em
\defaultaddspace=.5em
}

\newcommand{\kupf}{\ensuremath{\mathrm{Cu}_2\mathrm{O}}}

\renewcommand{\vec}[1]{\ensuremath{\mathbf{#1}}}

\begin{document}

\title{Exciton mass and exciton spectrum in the cuprous oxide}

\author{A. Alvermann}
\email{alvermann@physik.uni-greifswald.de}
\thanks{Author to whom any correspondence should be addressed.}
\author{H. Fehske}
\affiliation{Institut f\"ur Physik, Ernst-Moritz-Arndt-Universit\"at, 17487 Greifswald, Germany}

\begin{abstract}

Excitons with a radius of a few lattice constants can be affected by strong central-cell corrections,
leading to significant deviations of the optical spectrum from the hydrogen-like Rydberg series, and also to an enhancement of the exciton mass.
We present an approach to this situation based on a lattice model that incorporates the effects of a non-parabolic band structure, short distance corrections to the Coulomb interaction between electrons and holes, spin-orbit and exchange coupling.
The lattice model allows for observation of the crossover from large radius Wannier to small radius Frenkel excitons without invoking a continuum approximation.
We apply the lattice model approach especially to the yellow exciton series in the cuprous oxide,
for which the optical spectrum and exciton mass enhancement are obtained through adaptation of only a few model parameters to material-specific values.
Our results predict a strongly anisotropic ortho-exciton mass.
\end{abstract}

\maketitle

\section{Introduction}

Excitons in inorganic semiconductors such as silicon, germanium, or gallium arsenide
are usually well understood by Wannier theory, which describes excitons in analogy to the hydrogen atom~\cite{Knox63}.
Especially the exciton energies follow the Rydberg series $E_n = E_\mathrm{gap} - R_X/n^2$,
and the excitonic Rydberg $R_X$ can be computed from the dielectric constant and the reduced electron-hole mass.
In some semiconductors however,
with the cuprous oxide \kupf{} as one of the most prominent representatives~\cite{MS00,KFSSB14},
excitonic properties differ significantly from this simple picture.
Prominent features are the deviation of the energies of even exciton states from the Rydberg series,
and of the exciton mass from the sum of electron and hole mass.
These deviations can be attributed to central-cell corrections that become important whenever excitons are strongly bound states and thus sufficiently small to be affected by such corrections~\cite{KCB97}.

In the cuprous oxide, odd exciton states follow the Rydberg series rather accurately,
with an excitonic Rydberg $R_X\approx 97\, \mathrm{meV}$.
Even exciton states, on the other hand, defy Wannier theory:
The binding energy of the 1S (ortho-) exciton state is about $40\,\%$ larger than $R_X$, and the exciton mass is about $50\,\%$ larger than the sum of electron and hole mass (cf. Tab.~\ref{tab:CUO_X}).
Much theoretical effort has been invested into the exploration of this situation and its consequences~\cite{FKUS79,UFK81,THKAFBSG12,SMFWU16,SKGSSAHTFB16}.

\begin{table}[t]
\begin{tabular}{rcc}
\toprule\noalign{\vskip1ex}
 &  experiment &  \\\noalign{\vskip0.2ex}\cmidrule(lr){2-2}
\rule{0pt}{3.6ex} 
dielectric const.  \\
 static $\epsilon_s$ & 7.5   \\
 high freq. $\epsilon_\infty$ & 7.0 \\[0.2ex]
lattice constant $a$ & $0.42696 \, \mathrm{nm}$ \\[1ex]
electron mass $m_e$ &  $0.99 \, m_0$ & \\
hole mass $m_h$ ($\Gamma^+_7$) & $ 0.58 \, m_0$ & \\
band gap $E_\mathrm{gap}$ & $2.172\,\mathrm{eV}$ \\[1ex]
 &  & Wannier theory \\\noalign{\vskip0.2ex}\cmidrule{3-3}
\rule{0pt}{3.6ex} exciton mass $m_X$ & \hspace{14pt} $\approx 2.6 \, m_0$ (ortho)  & 
$1.7 \, m_0$ \\
& $(2.2$--$2.7) \, m_0$ (para)
\\[0.35ex]
excitonic Rydberg $R_X$ &  \multicolumn{1}{l}{$139 \, \mathrm{meV}$ (ortho)} & $97 \, \mathrm{meV}$ \\
 & \multicolumn{1}{l}{$151 \, \mathrm{meV}$ (para)} \\[0.25ex]
exc. Bohr radius $a_X$ (1S) & --- & $  8 \text{\AA} \approx 1.8 a$ \\
\noalign{\vskip1ex}\bottomrule
\end{tabular}
\caption{Experimental data relevant for yellow excitons in the cuprous oxide $\mathrm{Cu}_2\mathrm{O}$
(cf. App. A in Ref.~\cite{MS00} and references cited therein, as well as the more recent Refs.~\cite{JFKB05,BFSBSN07,NASK12}).
Where meaningful, we compare to the predictions of Wannier theory.
The excitonic Bohr radius is determined from the experimental Rydberg energy.
Note that the experimental value for the hole mass is problematic~\cite{NASK12}, probably as a consequence of the strong non-parabolicity of the $\Gamma_7^+$ valence band (see Fig.~\ref{fig:Cuprous}).
}
\label{tab:CUO_X}
\end{table}

\begin{figure}
\hspace*{\fill}
\includegraphics[scale=0.21]{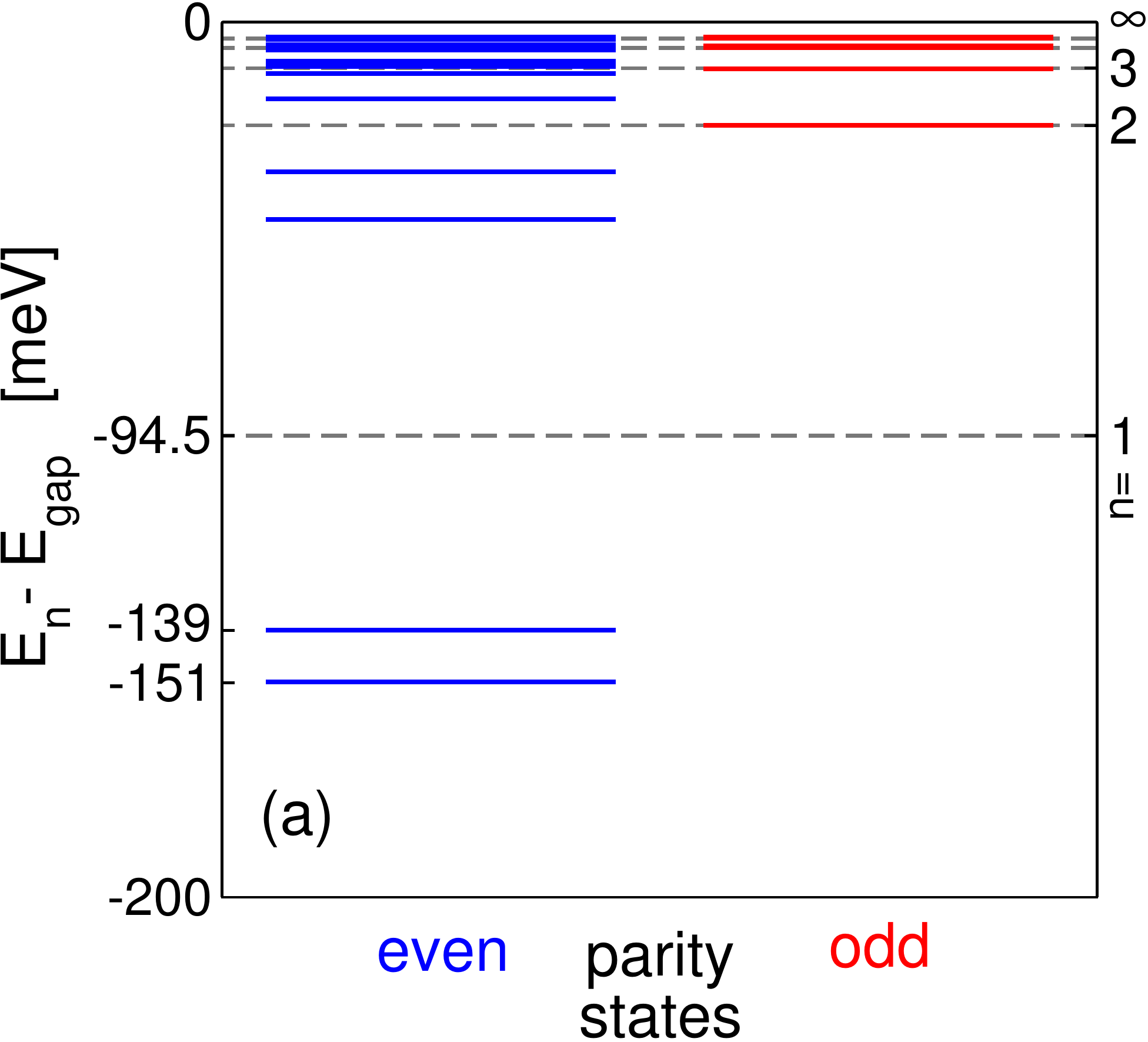}
\hspace*{\fill}
\includegraphics[scale=0.21]{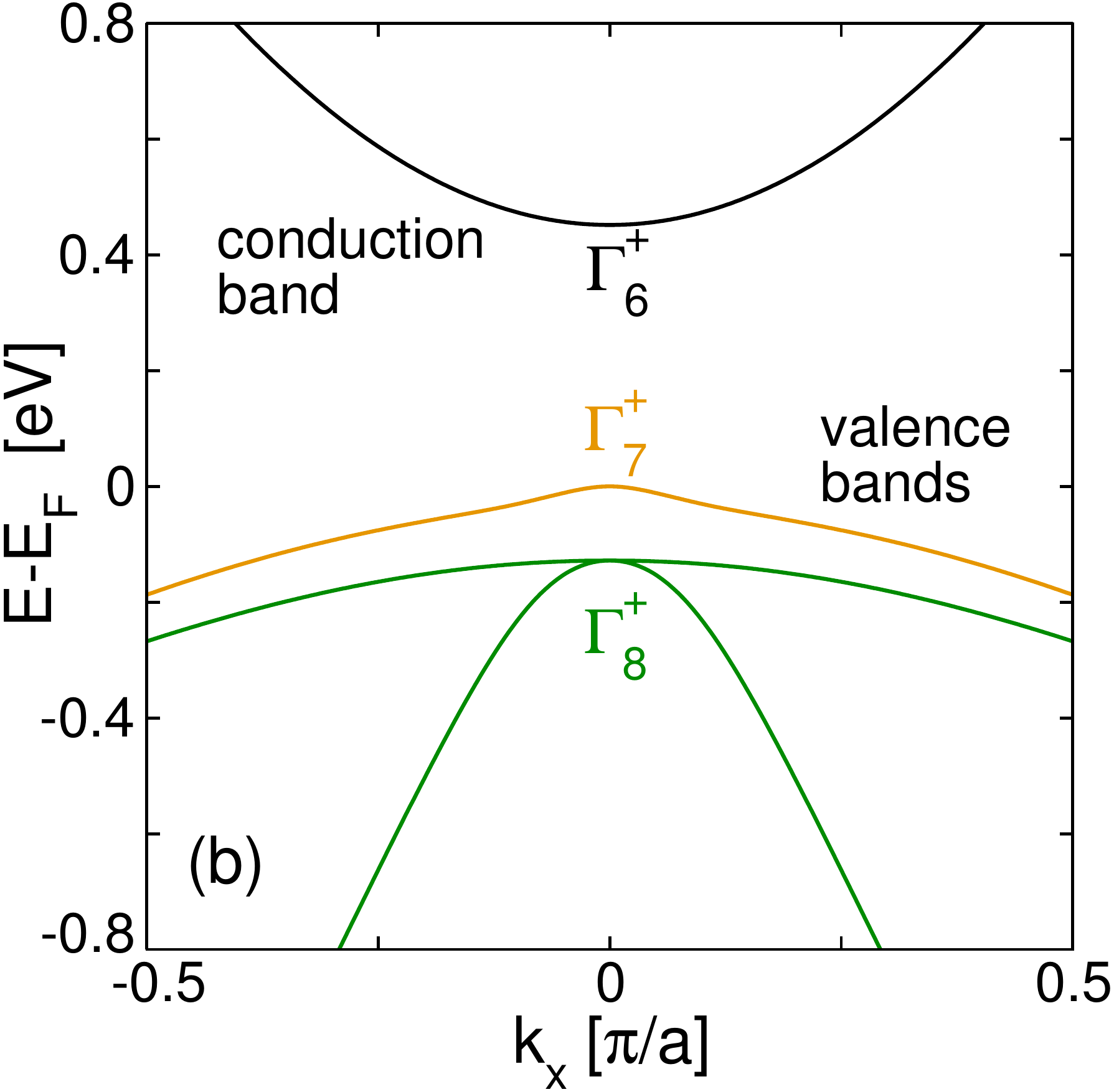}
\hspace*{\fill}
\caption{Left panel (a): Energies of even and odd exciton states in the cuprous oxide \kupf{},
according to the experimental data collected in Ref.~\cite{SMWU17}.
Dashed lines give the Rydberg series $E_n - E_\mathrm{gap} = -R_X / n^2$ of yellow excitons for comparison.
Right panel (b): Conduction and valence bands for the yellow and green exciton series in \kupf{}, according to the DFT data from Ref.~\cite{FSSR09}.
 }
\label{fig:Cuprous}
\end{figure}

One aspect of the explanation is hinted at by the non-parabolic dispersion of the upper valence band (see Fig.~\ref{fig:Cuprous}), which gives rise to the yellow exciton series. We should expect that such a strong violation of the effective mass approximation already at small momentum will lead to strong deviations from Wannier theory.

Recently, a comprehensive treatment of the even exciton spectrum
has been presented in Ref.~\cite{SMWU17},  with good agreement between theory and experiment.
This reference present a detailed analysis of spin-orbit coupling, exchange interaction, and central-cell corrections to the Coulomb attraction between electron and hole,
which are incorporated into an extended Luttinger-type~\cite{Lutt56,BL71,BL73,BL74,SH74,LA77,AL77} Hamiltonian. 

In this paper, we approach the problem of excitons with strong central-cell corrections from a different angle.
Our goal is to relate two disparate effects, the shift of the even exciton states relative to the Rydberg series and the enhancement of the exciton mass relative to the sum of hole and electron mass, through one theoretical description.
We will try to achieve this goal with a microscopic approach that is based on a lattice model instead of the continuum approximation.
Central-cell corrections occur naturally in the lattice model when excitons become small.

In the remainder of this paper we introduce the exciton lattice model in Sec.~\ref{sec:LKM}
and discuss the various terms, especially the relevance of spin-orbit coupling for the hole dispersion.
In Sec.~\ref{sec:X} we detail the numerical solution of the model and study the crossover from large to small excitons.
In Sec.~\ref{sec:CUO} we apply the theory to the cuprous oxide,
before we conclude in Sec.~\ref{sec:Conc}.

\section{Exciton lattice model}
\label{sec:LKM}

The exciton lattice model 
comprises three valence bands and one conduction band with cubic symmetry.
Its construction follows the general reasoning for Luttinger-type exciton models~\cite{SMWU17,Lutt56,BL71,BL73,BL74,SH74,LA77,AL77},
but in contrast to previous studies we consider the lattice structure explicitly  and do not invoke a continuum or effective mass approximation.

The valence bands are formed out of three orbitals $|\breve x\rangle, |\breve y \rangle, |\breve z \rangle$ at each lattice site, which transform according to the $\Gamma_5^+$ representation of the cubic symmetry group $O_h$~\cite{KDWS63}. We can picture these orbitals as d-type orbitals, associating $|\breve x \rangle  \simeq |d_{yz}\rangle$ etc.
The choice of the $\Gamma_5^+$ representation is specific to the cuprous oxide~\cite{Ell61,DS66}, 
other representations would be equally possible.
The hole states $|\vec r_h, \breve{ \mathsf d}, \mathsf s_h \rangle = |\vec r_h\rangle \otimes |\breve{\mathsf d} \rangle \otimes |\mathsf s_h \rangle$ are labelled by the hole position $\vec r_h$, orbital index $\mathsf d = x, y, z$, and hole spin orientation $\mathsf s_h = \uparrow, \downarrow$.

The conduction band is formed out of  a single orbital per lattice site,
which transforms according to the $\Gamma_1^+$ representation of $O_h$.
The electron states $|\vec r_e, \mathsf s_e \rangle = |\vec r_e\rangle \otimes |\mathsf s_e\rangle$ are labelled
by the electron position $\vec r_e$ and spin orientation $\mathsf s_e = \uparrow, \downarrow$.

In combination, the hole--electron states $| \vec r_h, \mathsf d, \mathsf s_h\rangle \otimes | \vec r_e, \mathsf s_e \rangle$ that appear in the excitonic wave function have five
indices.
The entries of the hole and electron position vectors $\vec r_h$, $\vec r_e$ are integer multiples of the lattice constant $a$, the remaining indices $\mathsf d, \mathsf s_h, \mathsf s_e$ can assume $3 \times 2 \times 2$ different values.

The five terms in the lattice model Hamiltonian
\begin{equation}\label{HX}
  H_X = - H_\mathrm{hole} - H_\mathrm{so} + H_\mathrm{electron} + H_\mathrm{Coulomb}
   + H_\mathrm{ex} 
\end{equation}
capture the most important contributions to exciton formation:
The hole ($H_\mathrm{hole}$) and electron ($H_\mathrm{electron}$) kinetic energy and
the Coulomb attraction between electron and hole ($H_\mathrm{Coulomb}$).
The spin-orbit coupling for the hole states ($H_\mathrm{so}$) is instrumental for explaining the peculiar shape of the valence bands, and thus of the yellow excitons in the cuprous oxide.
The exchange interaction ($H_\mathrm{ex}$), which depends on the relative orientation of electron and hole spin, leads to the splitting of ortho (triplet) and para (singlet) excitons.

For the present study we restrict ourselves to the most important leading order terms,
and thus minimize the number of adaptable model parameters.
For a fully accurate description of excitonic properties, extension of the model by additional terms might be necessary.

\subsection{Hole kinetic energy}

The leading term in the hole kinetic energy is the nearest-neighbor hopping term
\begin{equation}\label{HHole}
 H_\mathrm{hole} =  \sum_{\mathsf d = x,y,z}  \big(t_1 (1-I_\mathsf{d}^2) + t_2 I_\mathsf{d}^2\big) \big(T(\vec e_\mathsf{d}) + T(-\vec e_\mathsf{d}) \big) \;.
 \end{equation}
In this expression, $T(\vec e)$ denotes the translation operator in direction $\vec e$, i.e., $T(\vec e) |\vec r_h\rangle = |\vec r_h +  a \vec e \rangle$ with the lattice constant $a$.
The vector of nearest neighbour translations is denoted by $\vec e_\mathsf{d}$, i.e.,
$\vec e_x = (1,0,0)^t$ etc.

The operators $I_{\mathsf d}$ for $\mathsf d = x,y,z$ act on the hole states according to
\begin{alignat}{3}
I_x |\breve x \rangle &= 0 \;,
&
\quad
 I_y |\breve x \rangle &= - \ii |\breve z\rangle \;,
 &
\quad
 I_z |\breve x \rangle &= - \ii |\breve y\rangle \;,
\nonumber \\[1ex]
 I_x |\breve y \rangle &= - \ii |\breve z\rangle \;,  
&
 I_y |\breve y \rangle &= 0 \;,
 &
 I_z |\breve y \rangle &= + \ii |\breve x\rangle  \;, 
 \nonumber \\[1ex]
 I_x |\breve z \rangle &= + \ii |\breve y\rangle \;,
 &
 I_y |\breve z \rangle &= + \ii |\breve x\rangle \;,
 &
 I_z |\breve z \rangle &= 0 \;.
 \end{alignat}
The $I_{\mathsf d}$ fulfil, apart from a factor $\hbar$, angular momentum commutation relations, $[I_x, I_y] = \ii I_z$ etc.
The operator $1- I_{\mathsf d}^2$ is the projection onto orbital $|\breve{\mathsf d} \rangle$,
and $I_{\mathsf d}^2$ the projection onto the two other orbitals.

Compatibility of the kinetic energy term~\eqref{HHole} with the $\Gamma_5^+$ symmetry of the valence band orbitals is obvious.
Higher order terms are restricted by symmetry considerations,
and given in App.~\ref{app:Kinetic}.
Note that the leading term~\eqref{HHole} is equally valid for $\Gamma_5^+$ (d-type) and $\Gamma_5^-$ (p-type) symmetry of the valence band orbitals, but subsequent terms differ. 

\begin{figure}
\hspace*{\fill}
\includegraphics[scale=0.21]{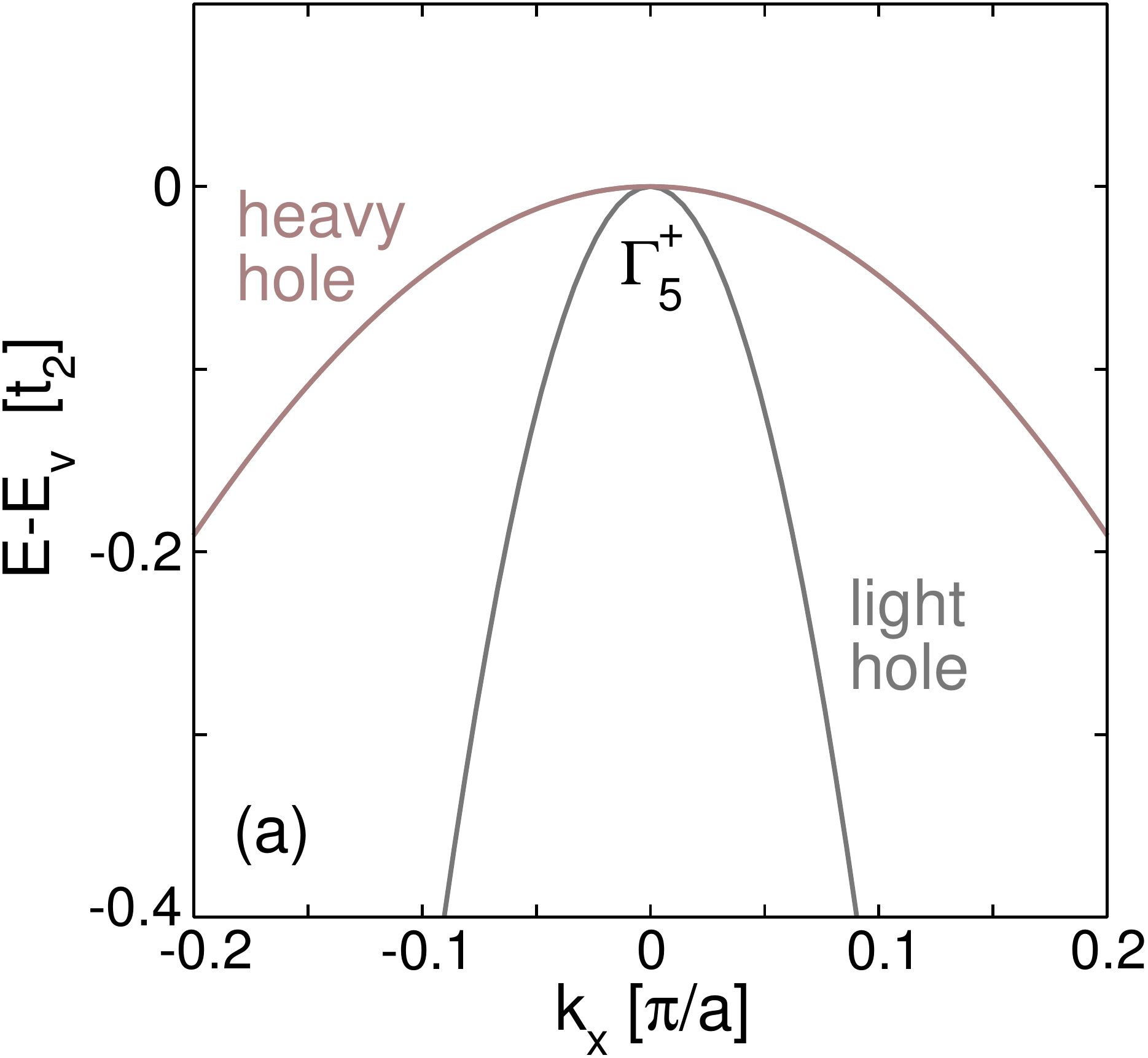}
\hspace*{\fill}
\includegraphics[scale=0.21]{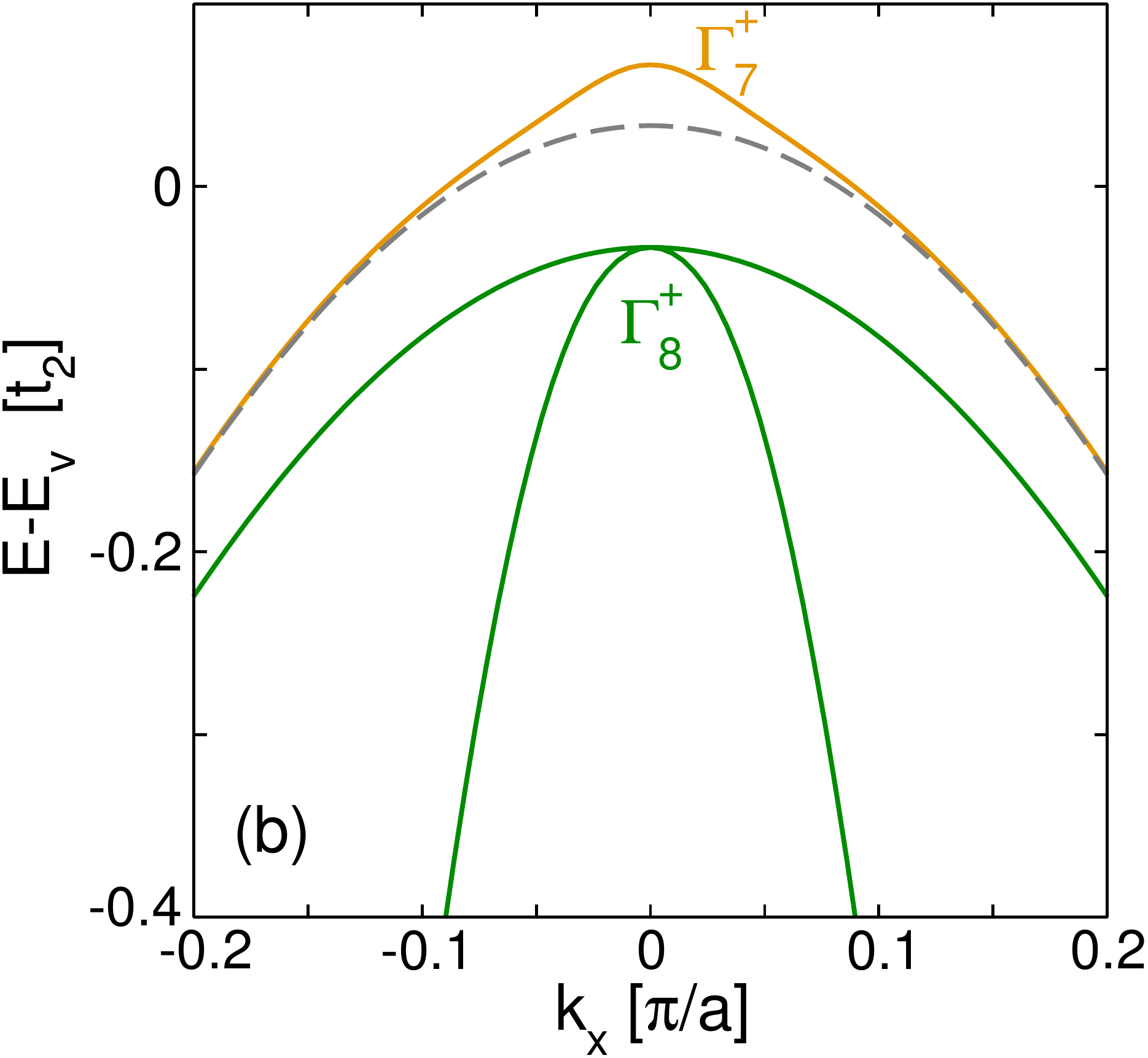}
\hspace*{\fill}
\caption{Hole dispersion without (left panel) and with (right panel) spin-orbit coupling.
The left panel (a) shows the valence bands according to $H_\mathrm{hole}$,
the right panel (b) includes spin-orbit coupling according to $H_\mathrm{so}$.
Parameters are $t_1/t_2 = 10$, $E_\mathrm{so}/t_2 = 0.1$.
}
\label{fig:valence1}
\end{figure}

The hole kinetic energy~\eqref{HHole} describes three valence bands $\mathsf d = x,y,z$ (see Fig.~\ref{fig:valence1})
with dispersion
\begin{equation}
E_h^{(\mathsf d)} (\vec k_h) =  2 t_1 \cos a (\vec k_h \cdot \vec e_{\mathsf d})  + 2 t_2 \sum_{i \ne \mathsf d} \cos a ( \vec k_h \cdot \vec e_i)  \;.
\end{equation}
On physical grounds, considering the overlap of valence band orbitals with orientations parallel and perpendicular to the direction of hole momentum $\vec k_h$, we expect $0 < t_2 < t_1$.
The corresponding light ($m_\mathrm{lh}$) and heavy ($m_\mathrm{hh}$) hole mass, along the direction of a lattice axis $\vec k_h \parallel \vec e_\mathsf{d}$, is
$m_\mathrm{lh} = \hbar^2/(2 a^2 t_1)$ and $m_\mathrm{hh} = \hbar^2/(2 a^2 t_2)$.
In intermediate directions, the hole mass interpolates between these two values.

\subsection{Spin-orbit coupling}
The spin-orbit coupling term for the hole takes the form
\begin{equation}\label{HSO}
 H_\mathrm{so} = - \frac{2 E_\mathrm{so} }{3 \hbar} \, (\vec I \cdot \vec S^h) \;,
\end{equation}
with the hole spin operator $\vec S^h$.
It connects hole states within the same unit cell.
The minus sign in the spin-orbit coupling term is explained effortlessly by the d-type nature of the valence band orbitals.

At $\vec k_h=0$ the hole states possess full cubic symmetry,
which results in a four-fold and a two-fold degenerate state according to the splitting
$\Gamma_5^+ \otimes \Gamma_6^+ = \Gamma_7^+ + \Gamma_8^+$ of the orbital-spin representation for valence band states~\cite{KDWS63}.
For general $\vec k_h \ne 0$ the symmetry is reduced, and only a two-fold degeneracy remains.
The two-fold degeneracy is associated with reflection symmetry of the valence band Hamiltonian $H_\mathrm{hole} + H_\mathrm{so}$ (see App.~\ref{app:HoleBands}).

Because of the minus sign in Eq.~\eqref{HSO}, the two-fold degenerate ``split-off'' band is shifted to higher energies.
Yellow excitons in the cuprous oxide involve holes in this split-off band.
At $\vec k_h = 0$, the two set of states are separated by the spin-orbit coupling energy $E_\mathrm{so}$,
as depicted in Fig.~\ref{fig:valence1}.

The full hole dispersion can be obtained through diagonalization of a $3 \times 3$ matrix (see App.~\ref{app:HoleBands}).
For the split-off band in Fig.~\ref{fig:valence1}b (right panel) we have the following situation.
For small $|\vec k| \to 0$, the dispersion along the direction of a lattice axis, e.g., along the $[100]$ direction $\vec e_x$, is
\begin{multline}
 E(k_x) - E_v = \frac23 E_\mathrm{so} - \frac13 (t_1 + 2 t_2) (a k_x)^2 + \\ \frac{8(t_1-t_2)^2 + E_\mathrm{so} (t_1 + 2 t_2) }{36 E_\mathrm{so}} (a k_x)^4 + O \big( (a k_x)^6 \big) \;,
\end{multline}
which gives the new hole mass in  the split-off band
\begin{equation}\label{mso}
m_\mathrm{s-o} =\frac{3 \hbar^2}{2 a^2 (t_1 + 2 t_2)} = \frac{3  m_\mathrm{lh}  m_\mathrm{hh}}{2  m_\mathrm{lh} +  m_\mathrm{hh}} \;,
\end{equation}
 and a pronounced non-parabolicity from the $k^4$ term.
Here, $E_v = 2 t_1 + 4 t_2$ is the energy of the valence bands at $\vec k = 0$ according to $H_\mathrm{hole}$, i.e., essentially the Fermi energy  in the absence of spin-orbit coupling.
We have $m_\mathrm{lh} < m_\mathrm{s-o} < m_\mathrm{hh}$.

For larger $|\vec k|$, in the region of $ E_\mathrm{so}  \lesssim 2 (t_1 - t_2) (a k_x)^2$,
the dispersion changes into
\begin{equation}
  E(k_x) - E_v =  \frac13 E_\mathrm{so}  + 2 t_2 (\cos a k_x - 1) \;,
\end{equation}
which is the dispersion of the heavy hole band with mass $m_\mathrm{hh}$, shifted by spin-orbit coupling.
As a consequence of spin-orbit coupling, the hole mass in the split-off band changes significantly 
already for small $|\vec k|$. 
The pronounced non-parabolicity of the hole dispersion is partly responsible for the peculiar properties of yellow excitons in the cuprous oxide.

\subsection{Electron kinetic energy}

For one conduction band, the leading term of the electron kinetic energy has the simple form
\begin{equation}\label{HE}
 H_\mathrm{electron} = - t_e \sum_{i = x,y,z} (T(\vec e_i) + T(-\vec e_i) ) \;,
\end{equation}
where $T(\vec e)$ is still the translation operator with $T(\vec e) |\vec r_e \rangle = |\vec r_e + a \vec e\rangle$.
The Hamiltonian~\eqref{HE} describes a conduction band with cosine dispersion $E_e(\vec k_e) = - 2 t_e \sum_i \cos a (\vec k_e \cdot \vec e_i)$ and electron mass $m_e = \hbar^2/(2 a^2 t_e)$.
Subsequent terms in the electron kinetic energy are restricted by cubic symmetry,
and can be assembled in analogy to
the hole kinetic energy terms (cf. App.~\ref{app:Kinetic}).

\subsection{Coulomb attraction}

Excitons form through the Coulomb attraction between electron and hole.
We assume an interaction
\begin{equation}
 H_\mathrm{Coulomb} =  \sum\limits_{\vec r_h, \, \vec r_e} U(\vec r_e - \vec r_h)  |\vec r_h\rangle\langle \vec r_h| \otimes   |\vec r_e\rangle\langle \vec r_e|  \;,
\end{equation}
which is diagonal in the spin, orbital, and lattice indices,
and choose the expression
\begin{equation}
 U(\vec r) = -  e^2 \times \begin{cases} \dfrac{1}{\epsilon |\vec r|} \quad & \text{ if } \vec r \ne 0 \;, \\[3ex]
 \;\dfrac{1}{\ell_C} & \text{ if } \vec r = 0 \end{cases}
\end{equation}
for the dependence on the electron-hole separation.
Here, $e$ is the elementary charge and $\epsilon$ a material-specific dielectric constant.
The parameter $\ell_C$ is a characteristic length comparable to the width of the valence and conduction band orbitals.

For large electron-hole separation, the Coulomb potential decays $\propto 1/| \vec r_e - \vec r_h |$,
while for $\vec r_e = \vec r_h$ it assumes a finite value $U(0) \propto 1/\ell_C$.
Note that a finite local value $U(0)$ does not result from the screening of the Coulomb attraction,
but from the finite size of the valence and conduction band orbitals.

The correct choice of the dielectric constant $\epsilon$ is debatable,
as screening of the Coulomb attraction depends on distance.
Interpolation between the different values of the dielectric constant is possible through modified expressions for the Coulomb potential (see, e.g., Ref.~\cite{PB77}).
To avoid 
 the introduction of additional parameters,
we keep the above simpler expression. Some of the distance dependence of screening can be captured through adaptation of the model parameter $\ell_C$.

\subsection{Exchange interaction}

Exchange interaction leads to the splitting of exciton states into ortho--(triplet)--excitons and para--(singlet--) excitons.
We assume a local exchange interaction
\begin{equation}
 H_\mathrm{ex} = E_\mathrm{ex} \sum_{\vec r_h = \vec r_e} \left(\frac{1}{4} - \frac{1}{\hbar^2} \vec S^h \cdot \vec S^e \right) \otimes  |\vec r_h \rangle \langle \vec r_h| \otimes |\vec r_e \rangle \langle \vec r_e | \;,
\end{equation}
with the hole (electron) spin operator $\vec S^h$ ($\vec S^e$),
which connects electron and hole states in the same unit cell and is diagonal in the orbital indices.
In the limit of zero exciton radius, the 1S ortho--exciton state is shifted by $E_\mathrm{ex}$ relative to the 1S para--exciton state, which is not affected.
The level splitting decreases with increasing exciton radius, and is zero for odd exciton states.

\section{Excitons in the lattice model}
\label{sec:X}

Our computations determine the 
exciton wave function
\begin{equation}
 |\psi_{\vec K}\rangle = \sum\limits_{\vec r_e} e^{\ii \vec K \cdot \vec r_e} \big( T(\vec r_e) |\phi_{\vec K}\rangle \big) \otimes |\vec r_e \rangle \;,
\end{equation}
which is the translational invariant extension
of the relative hole-electron wave function
\begin{equation}
 |\phi_{\vec K} \rangle =  \sum\limits_{\vec r_h} \sum_{\mathsf d = x, y, z} \sum_{\mathsf s_h, \mathsf s_e = \uparrow, \downarrow} \phi_{\vec K}(\vec r_h, \mathsf d, \mathsf s_h, \mathsf s_e) |\vec r_h, \breve{\mathsf d} , \mathsf s_h \rangle \otimes | \mathsf s_e \rangle \;.
\end{equation}
The latter wave function is an eigenstate of the Bloch Hamiltonian $H_X(\vec K)$, parametrized by the exciton momentum $\vec K$.
In the above expression for $|\phi_{\vec K}\rangle$, we fix the electron position $\vec r_e$ and use the hole position $\vec r_h$ as the relative coordinate.
Any other convention is equally possible, and we can swap the role of the electron and hole.

We compute the exciton wave function and energies numerically with standard sparse matrix eigenvalue solvers~\cite{Sa92,So02} applied to $H_X(\vec K)$.
The numerical effort can be reduced considerably by using an adaptive scheme to deal with the increasing radius of excited exciton states~\cite{ALF11}.

From the exciton wave function at zero momentum $\vec K =0$, the exciton radius is obtained as the expectation value
\begin{equation}
a_X = \frac23 \langle r \rangle = \frac23 \langle (x^2 + y^2 + z^2)^{1/2} \rangle \;.
\end{equation}
We include the prefactor $2/3$ to recover the Bohr radius, which is defined as the most probable radial distance, in the continuum limit where $\phi_{\vec K}(r) \propto e^{-r/a_X}$.

From the exciton energy $E(\vec K)$ at finite momentum, the exciton mass is obtained as the derivative
\begin{equation}
m_X^{-1} = \frac{1}{\hbar^2} \frac{ \partial^2 E(k \, \vec e_\mathsf{d})}{\partial k^2} \Big|_{k=0} \;.
\end{equation}
In general, the exciton mass is anisotropic, but we will only report its values along a lattice axis $\vec K \parallel \vec e_\mathsf{d}$.

Note that in order to obtain the exciton energies $E_n - E_\mathrm{gap}$ the computed eigenvalues of $H_X$ have to be shifted by a constant $E_\mathrm{gr} =  2 t_1 + 4 t_2 + 6  t_e + (2/3) E_\mathrm{so}$ ---the groundstate energy of $H_X$ at zero Coulomb attraction--- that is determined by the above form of the kinetic energy and spin-orbit coupling terms.
From the exciton energies, the spectral lines are obtained by adding the experimental band gap value $E_\mathrm{gap} = 2.172 \, \mathrm{eV}$, which is a free constant that does not enter the Hamiltonian $H_X$.

\subsection{Symmetry considerations}

Due to the coupling of the orbital and spin degrees of freedom and the reduced symmetry in the lattice model in comparison to the continuum description, most simple observables fail to be conserved.
In particular, neither the effective orbital spin $\vec I$, hole spin $\vec S^h$,
electron spin $\vec S^e$ nor any of their combinations commute with the full Hamiltonian.
Already for the hole Hamiltonian $H_\mathrm{hole} + H_\mathrm{so}$, the effective orbital-hole spin $\vec I + \vec S^h$ is conserved only at $\vec k_h = 0$ or other points of high symmetry.

The minimal symmetry that is preserved at least for zero exciton momentum $\vec K = 0$ is the cubic symmetry $O_h$.
We do not use the entire group $O_h$ to set up the computational problem,
but use the following easier construction that results in a comparable reduction  of the computational effort.

The Hamiltonian $H_X$ and the three operators
$R_\mathsf{d} = (2 I_\mathsf{d}^2 - 1) \otimes (2/\hbar) S^h_\mathsf{d} \otimes (2/\hbar) S^e_\mathsf{d}$
for $\mathsf d = x,y,z$ commute among themselves.
This allows us to split the Hamiltonian,
with its $3 \times 2 \times 2 = 12$ local degrees of freedom (orbital, hole spin, electron spin),
into four operators acting on three local degrees of freedom each.
The respective basis states are listed in App.~\ref{app:basis}.

Geometrically, the operators $R_\mathsf{d}$ implement a local reflection in orbital and spin space,
and thus provide a representation of the Abelian group $\mathbb Z_2 \times \mathbb Z_2 \times \mathbb Z_2 \simeq D_\mathrm{2h}$. 
The fact that $H_X$ and the $R_\mathsf{d}$ commute holds true only for the simple hole kinetic energy expression in Eq.~\eqref{HHole}.
Extension of the symmetries to a general hole kinetic energy beyond the leading term 
requires combination of the operators $R_d$ with a reflection in position space, e.g.,
$(x,y,z) \mapsto (-x,y,z)$ for $R_x$.
Here, we can consider all reflections independently. In this way, the computational effort can be reduced by a factor $1/(4 \times 2^3)=1/32$.

\begin{figure}
\hspace*{\fill}
\hspace*{2ex}\includegraphics[scale=0.27]{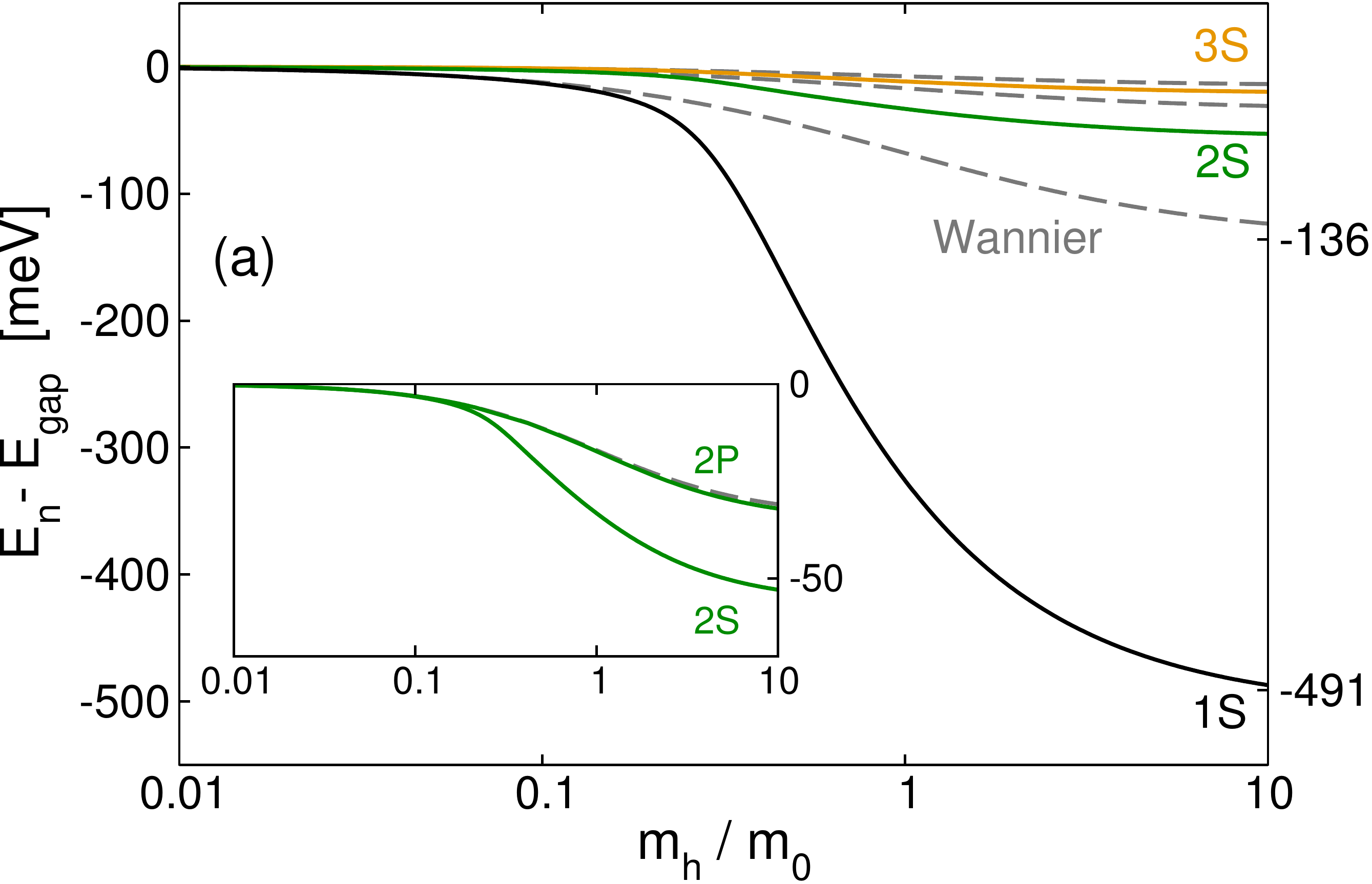} 
\hspace*{\fill}
\\[1ex]
\hspace*{\fill}
\includegraphics[scale=0.27]{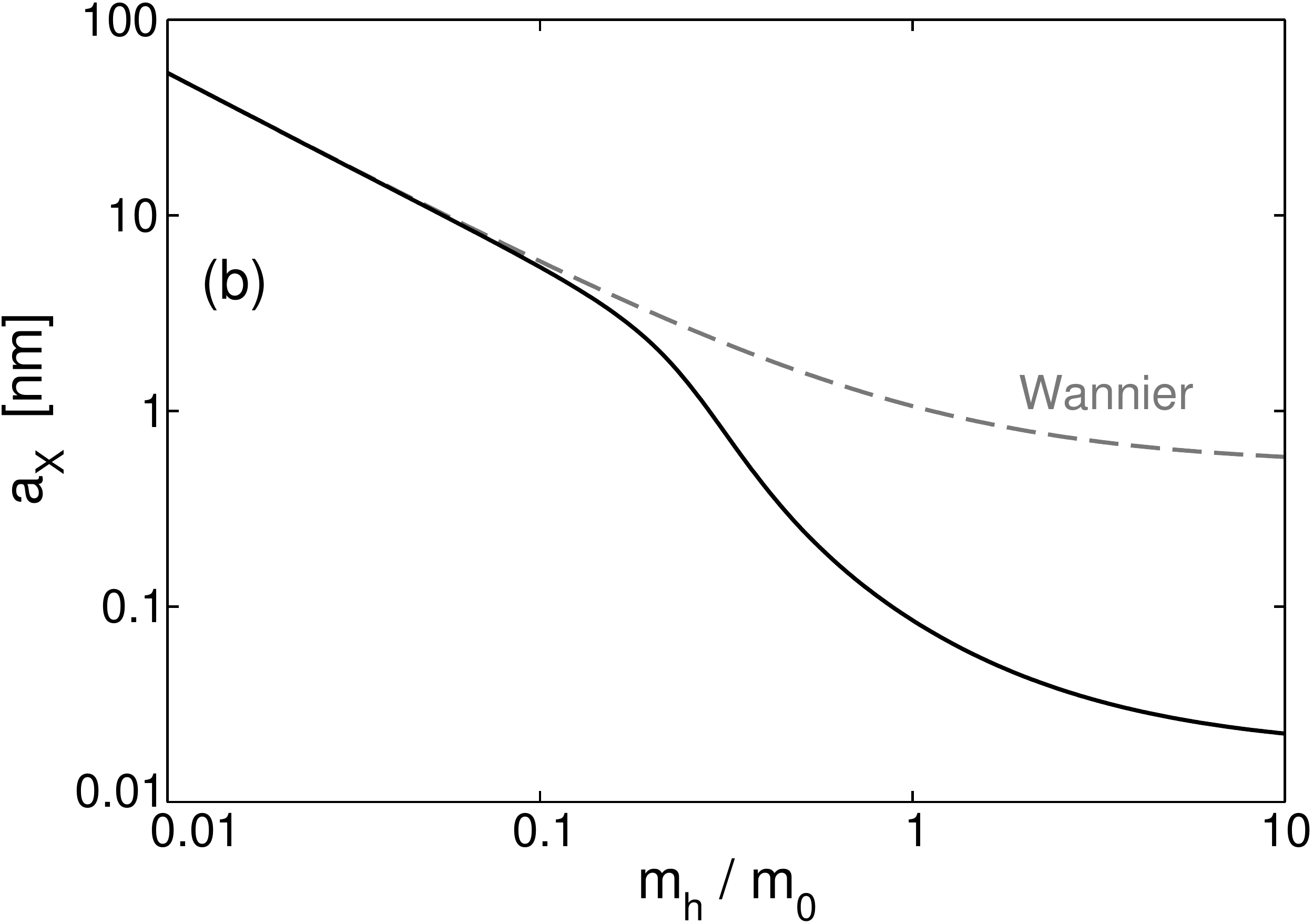} 
\hspace*{\fill}
\\[1ex]
\hspace*{\fill}
\hspace*{3pt}
\includegraphics[scale=0.27]{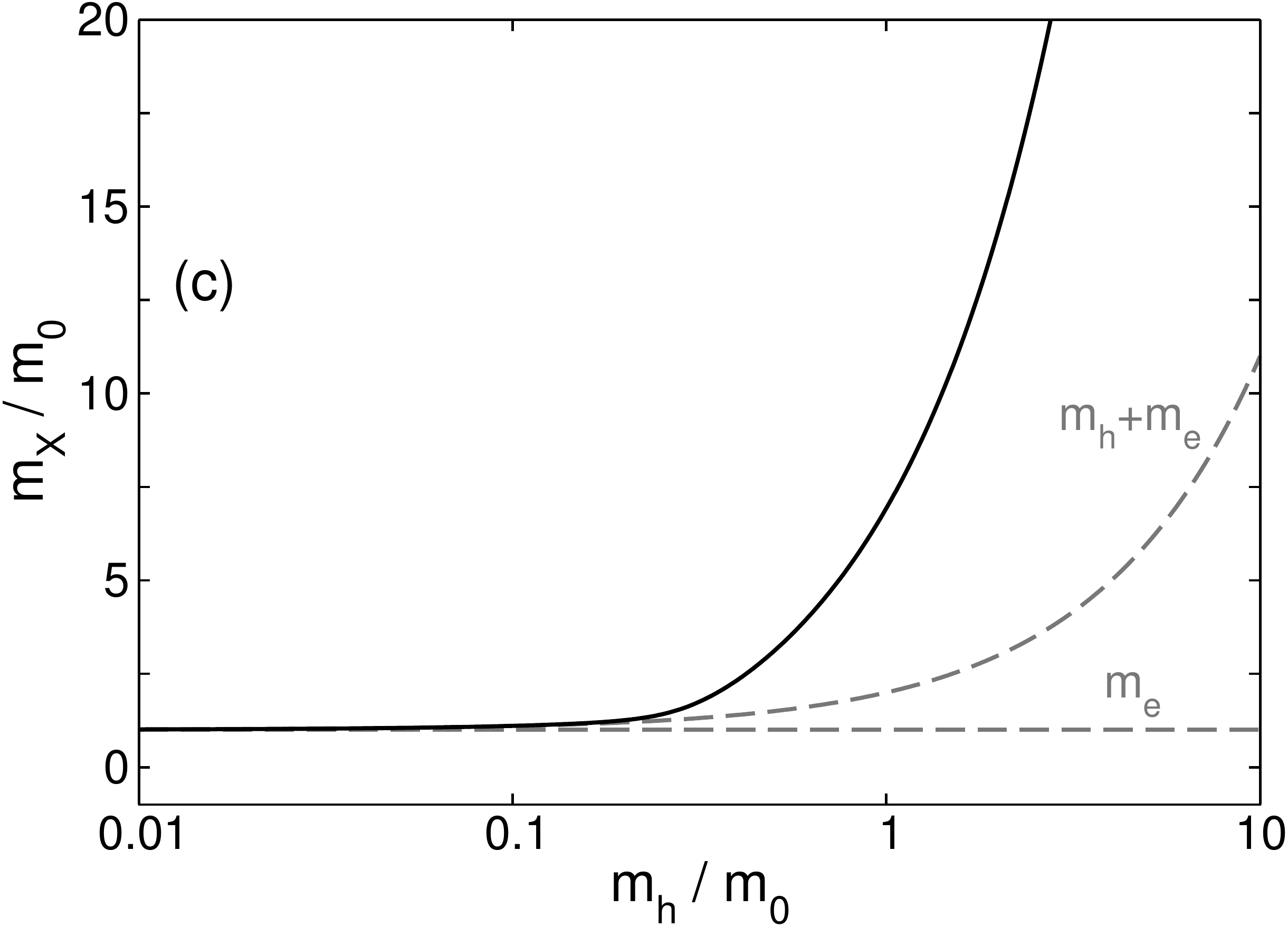}
\hspace*{\fill}
\caption{Crossover from large to small excitons, as a function of the hole mass $m_\mathrm{lh} = m_\mathrm{hh}\equiv m_h$.
Top panel (a): Energy of the first three even (1S, 2S, 3S) exciton states.
The inset gives the energy of the 2S and 2P states.
Middle panel (b): Exciton radius in the lowest (1S) state.
Bottom panel (c): Exciton mass $m_X$ in the lowest (1S) state.
In all panels, the dashed lines give the corresponding result obtained within the Wannier theory of hydrogen-like excitons.
}
\label{fig:Crossover}
\end{figure}

\subsection{Crossover from large to small excitons}

Formally, the crossover from large to small excitons can be achieved by scaling of the lattice constant $a$.
Physically, the crossover from large to small excitons takes place, for example,
when the hole mass increases.
In the Wannier theory of continuum excitons, 
the excitonic Rydberg (i.e., binding energy) $R_X$ and excitonic radius $a_X$ are given by
\begin{equation}
 R_X =  R_H \, \frac{\mu}{\epsilon^2} 
 \;, \quad
 a_X = a_B \, \frac{\epsilon}{\mu} \;,
 \end{equation}
with the Rydberg energy $R_H \approx 13.6 \, \mathrm{eV}$ of the hydrogen atom and
the Bohr radius $a_B \approx 0.053 \, \mathrm{nm}$.
These expressions involve only the reduced electron-hole mass $\mu = ( m_0/m_h + m_0/m_e )^{-1}$,
with the elementary electron mass $m_0$,
 and the dielectric constant $\epsilon$.
Excitons become large for $m_h \to 0$, when $a_X \to \infty$ and $R_X \to 0$. 

In Fig.~\ref{fig:Crossover}
we show the exciton energy, radius, and mass as a function of variable hole mass $m_\mathrm{lh} = m_\mathrm{hh} \equiv m_h$.
The remaining model parameters are
$m_e = m_0$, 
a large spin-orbit coupling $E_\mathrm{so} = 300\,\mathrm{meV}$ to separate the valence bands (since
$m_\mathrm{lh} = m_\mathrm{hh}$ the precise value is not relevant),
lattice constant $a = 1\,\mathrm{nm}$,
dielectric constant $\epsilon=10$
and Coulomb length $\ell_C = 2 a$,
and for the sake of simplicity no exchange interaction $E_\mathrm{ex} = 0$.

As was to be expected, significant deviations from Wannier theory occur as soon as $a_X \simeq a$,
which happens here for $m_h / m_0 \gtrsim 0.2$.
As the exciton becomes small, its binding energy becomes much larger than the Wannier Rydberg $R_X$.
We observe the shift of the energy of the even (1S, 2S, 3S) exciton states, and the splitting of the first excited (2S, 2P) state in the reduced (cubic instead of full rotational) symmetry that now applies~\cite{Cho76}.
Note that the odd exciton states are nearly unaffected by an energy shift (in the inset in Fig.~\ref{fig:Crossover}a the two curves for the 2P state and the Wannier theory result can barely be distinguished).
For $m_h \to \infty$, the Wannier Rydberg would converge to the value $R_H m_e / (m_0 \epsilon^2)$,
while the true binding energy approaches the (for given parameters larger) value $e^2 / \ell_C + 6 t_e$ (both values are marked in Fig.~\ref{fig:Crossover}.)
The crossover to small excitons is accompanied by a strong enhancement $m_X/(m_h + m_e) \gg 1$ of the exciton mass, as the exciton becomes increasingly immobile with decreasing radius.
This situation, when kinetic energy is suppressed in favour of Coulomb attraction, is the regime of small Frenkel excitons.

\section{Yellow and green excitons in the cuprous oxide}
\label{sec:CUO}

\begin{table}
\begin{tabular}{rlc}
\toprule
\noalign{\vskip1ex}
 &  DFT bands &  spectrum \\\noalign{\vskip1ex} \cmidrule(lr){1-3}\noalign{\vskip1ex}
bare hole mass $m_\mathrm{lh}$  & \hspace{1ex} $0.16 \, m_0$   \\
                           $t_1$ & \hspace{1ex} $1.31 \, \hspace{4.5pt} \mathrm{eV}$  \\[0.5ex]
bare hole mass $m_\mathrm{hh}$  & \hspace{1ex} $3.10 \, m_0$  \\
                          $t_2$ & \hspace{1ex} $0.067 \, \mathrm{eV}$ \\[0.5ex]
spin-orbit coup. $E_\mathrm{so}$ &  \hspace{1ex} $0.128 \, \mathrm{eV}  $ \\[0.75ex]
dielectric constant $\epsilon$  && $6.94$   \\
\rule{1ex}{0pt} Coulomb length $\ell_C/a$ && $1.75$  \\
exchange energy  $E_\mathrm{ex}$ &&  $666 \, \mathrm{meV}$ 
\\
\noalign{\vskip1ex}
\bottomrule
\end{tabular}
\caption{Model parameters for the cuprous oxide $\mathrm{Cu}_2\mathrm{O}$.
The lattice constant $a = 0.42696 \, \mathrm{nm}$ and electron mass $m_e = 0.99 \, m_0$ (such that $t_e = 0.21 \, \mathrm{eV}$ in $H_\mathrm{electron}$) are taken from the experimental data given in Tab.~\ref{tab:CUO_X}.
}
\label{tab:CUO_params}
\end{table}

\subsection{Model parameters}

\begin{figure}
\hspace*{\fill}
\includegraphics[width=0.7\linewidth]{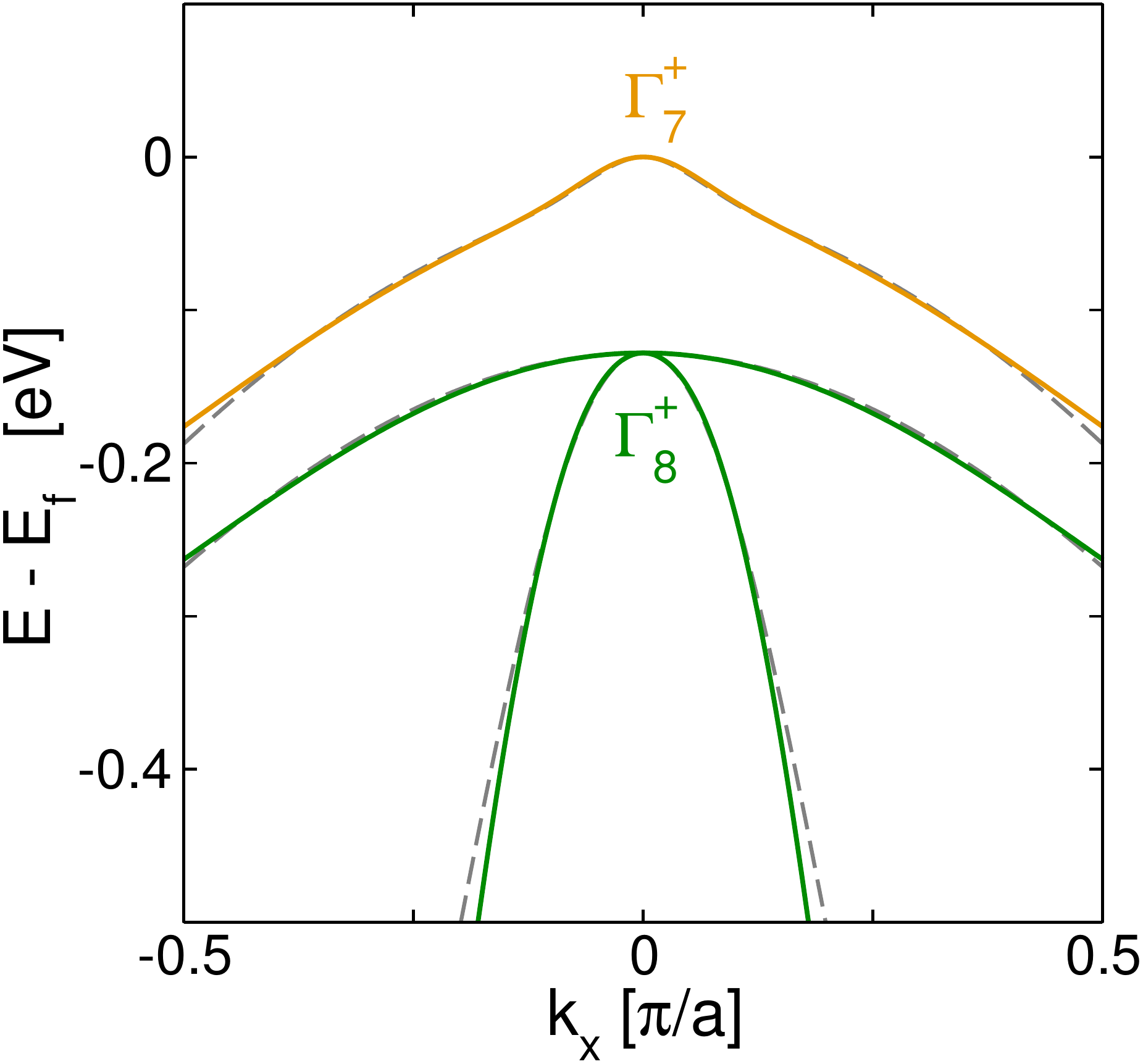}
\hspace*{\fill}
\caption{Fitting of  valence band parameters for the cuprous oxide \kupf{}.
Solid curves are the model bands from the Hamiltonian $H_\mathrm{hole} + H_\mathrm{so}$,
for parameters $m_\mathrm{lh} = 0.16 \, m_0$, $m_\mathrm{hh} = 3.1 \, m_0$, $E_\mathrm{so}=128\, \mathrm{meV}$ as in Tab.~\ref{tab:CUO_params}, grey dashed curves are the DFT bands from Fig.~\ref{fig:Cuprous}.
}
\label{fig:ValenceCuprous}
\end{figure}

The model Hamiltonian~\eqref{HX} depends on eight parameters.
Our logic for choosing these parameters is as follows:
(i) use the lattice constant $a$ and electron mass $m_e$ known from experiment;
(ii) trust the DFT valence bands, which fix the three parameters $m_\mathrm{lh}$, $m_\mathrm{hh}$, $E_\mathrm{so}$ of $H_\mathrm{hole} + H_\mathrm{so}$;
(iii) obtain the remaining three parameters $\epsilon$, $\ell_C$, $E_\mathrm{ex}$ from comparison with the experimental energies of the three lowest exciton states.

\begin{figure}
\hspace*{\fill}
\includegraphics[scale=0.27]{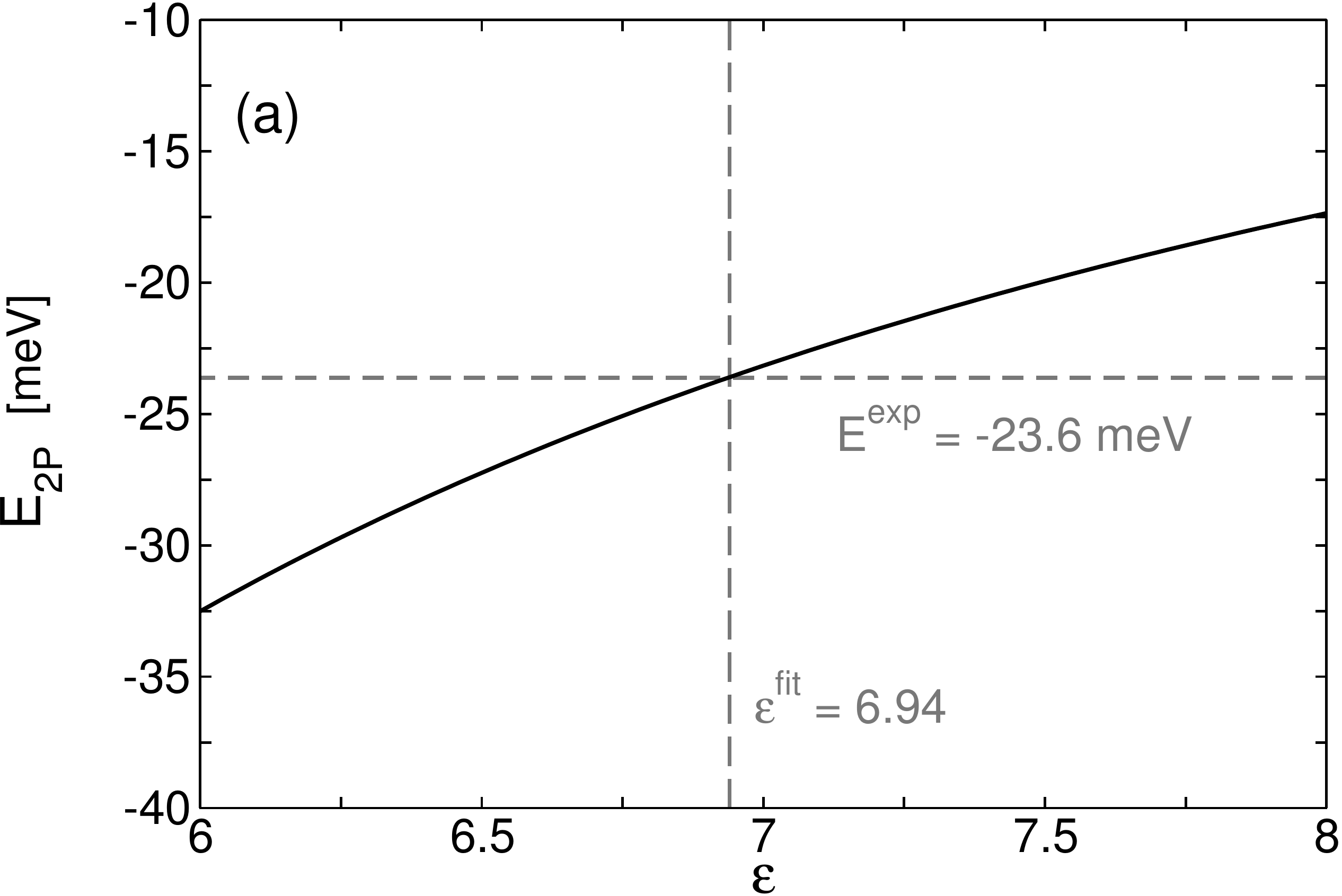}
\hspace*{\fill}
\\[1ex]
\hspace*{\fill}
\includegraphics[scale=0.27]{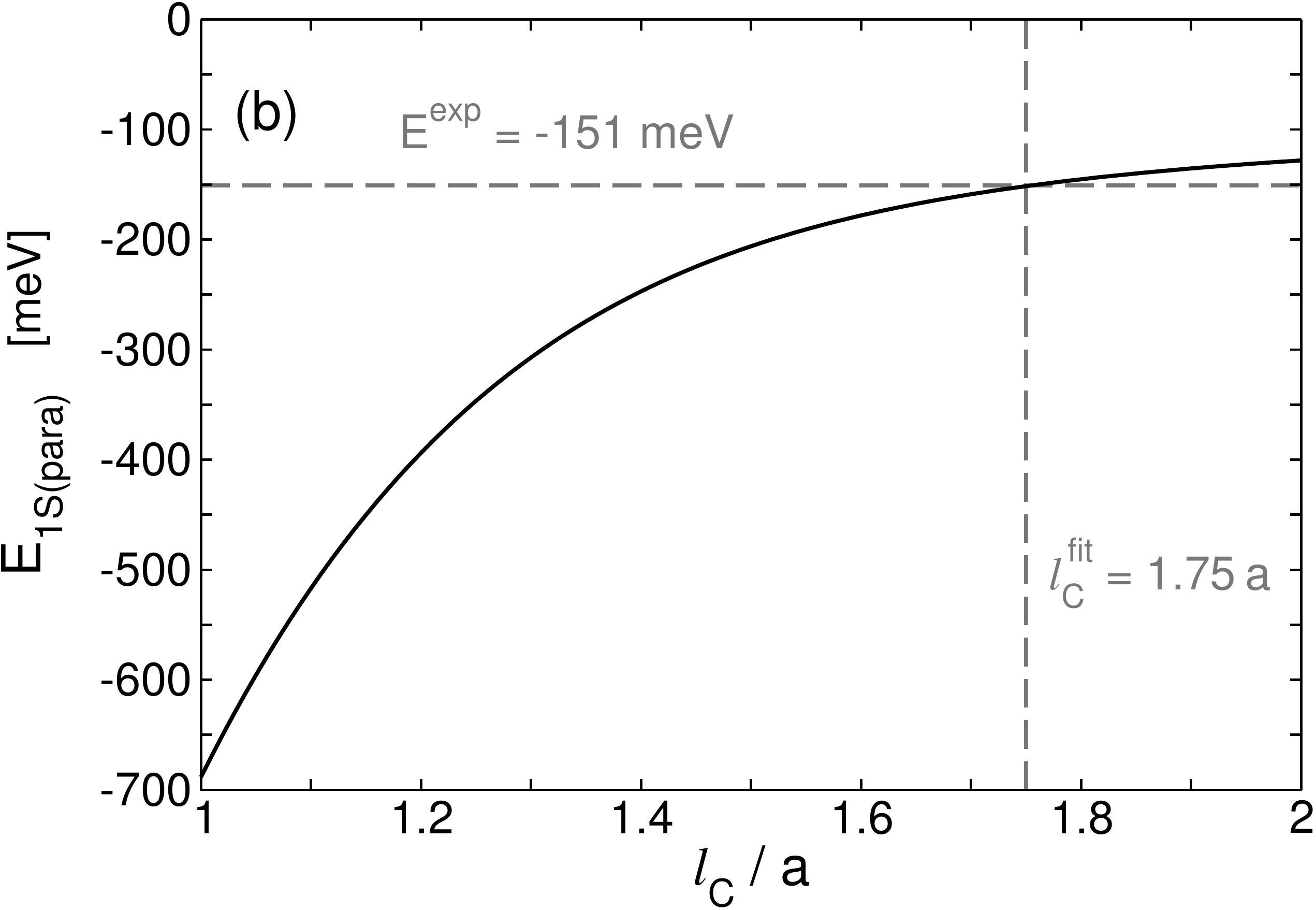}
\hspace*{\fill}
\\[1ex]
\hspace*{\fill}\hspace{6pt}
\includegraphics[scale=0.27]{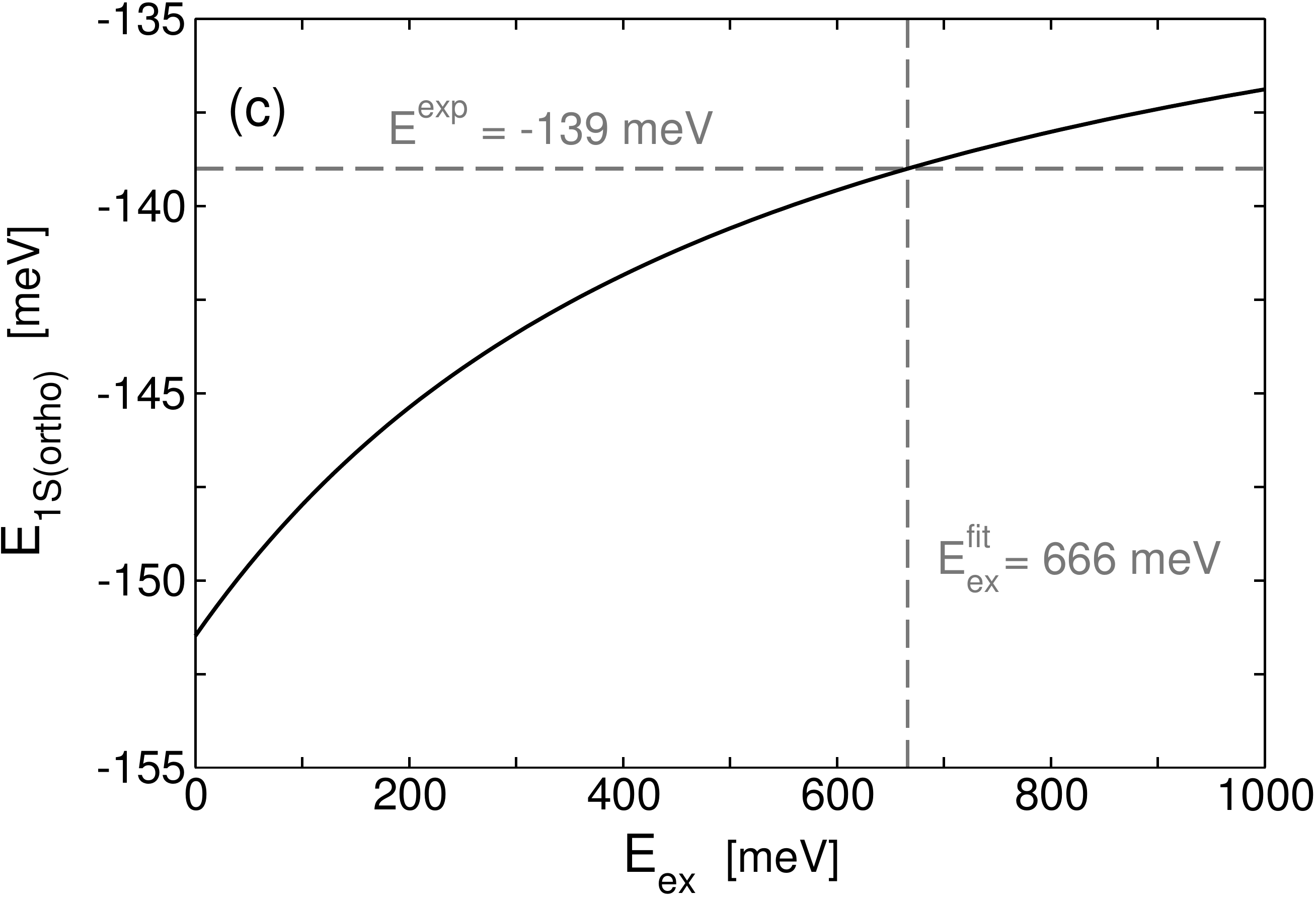}
\hspace*{\fill}
\caption{Fitting of model parameters for the cuprous oxide \kupf{}.
Top panel (a):
Energy of the $2P$ state as a function of the dielectric constant $\epsilon$.
Middle panel (b):
Energy of the 1S para-exciton state as a function of the Coulomb length $\ell_C$.
Bottom panel (c): Energy of the 1S ortho-exciton state as a function of the exchange energy $E_\mathrm{ex}$.
In all panels, the horizontal dashed lines give the experimental values,
and the vertical dashed lines the values of the model parameters resulting from the fit.
Energies are measured relative to the band gap $E_\mathrm{gap}$.
}
\label{fig:FitCuprous}
\end{figure}

The successive steps of parameter fitting are documented in Figs.~\ref{fig:ValenceCuprous},~\ref{fig:FitCuprous}, and lead to the model parameters in Tab.~\ref{tab:CUO_params}.
As seen in Fig.~\ref{fig:ValenceCuprous}, 
the DFT valence bands are very accurately reproduced at smaller values of $|\vec k|$ with the chosen parameters, including the non-parabolicity of the split-off band.
The DFT bands cannot be reproduced by the simple model kinetic energy at larger values of $|\vec k|$ (not shown), where the interaction with other bands become significant, but as we will see below these parts of the band structure are not relevant for exciton formation.
For a consistency check, we note that Eq.~\eqref{mso} gives $m_\mathrm{s-o} \simeq 0.44 m_0$ for the hole mass in the split-off band. At the present level of treatment, and with respect to the experimental uncertainties, this is consistent with the experimental hole mass $m_h = 0.58 m_0$ from Ref.~\cite{NASK12}.

In Fig.~\ref{fig:FitCuprous}, we first change $\epsilon$ and fit the energy of the 2P (ortho or para) exciton state, which does not depend on $\ell_C$ or $E_\mathrm{ex}$,
to the experimental value $E_\mathrm{2P}^\mathrm{exp} = -23.6\,\mathrm{meV}$.
Then, we change $\ell_C$ and fit the energy of the 1S para-exciton state,
which does not depend on $E_\mathrm{ex}$,
to the experimental value $E_\mathrm{1S(para)}^\mathrm{exp} = -151\,\mathrm{meV}$.
Finally, we change $E_\mathrm{ex}$ and fit the energy of the 1S ortho-exciton state
 to the experimental value $E_\mathrm{1S(ortho)}^\mathrm{exp} = -139\,\mathrm{meV}$.
Now, all model parameters have been determined.

\begin{figure}
\hspace*{\fill}
\includegraphics[scale=0.23]{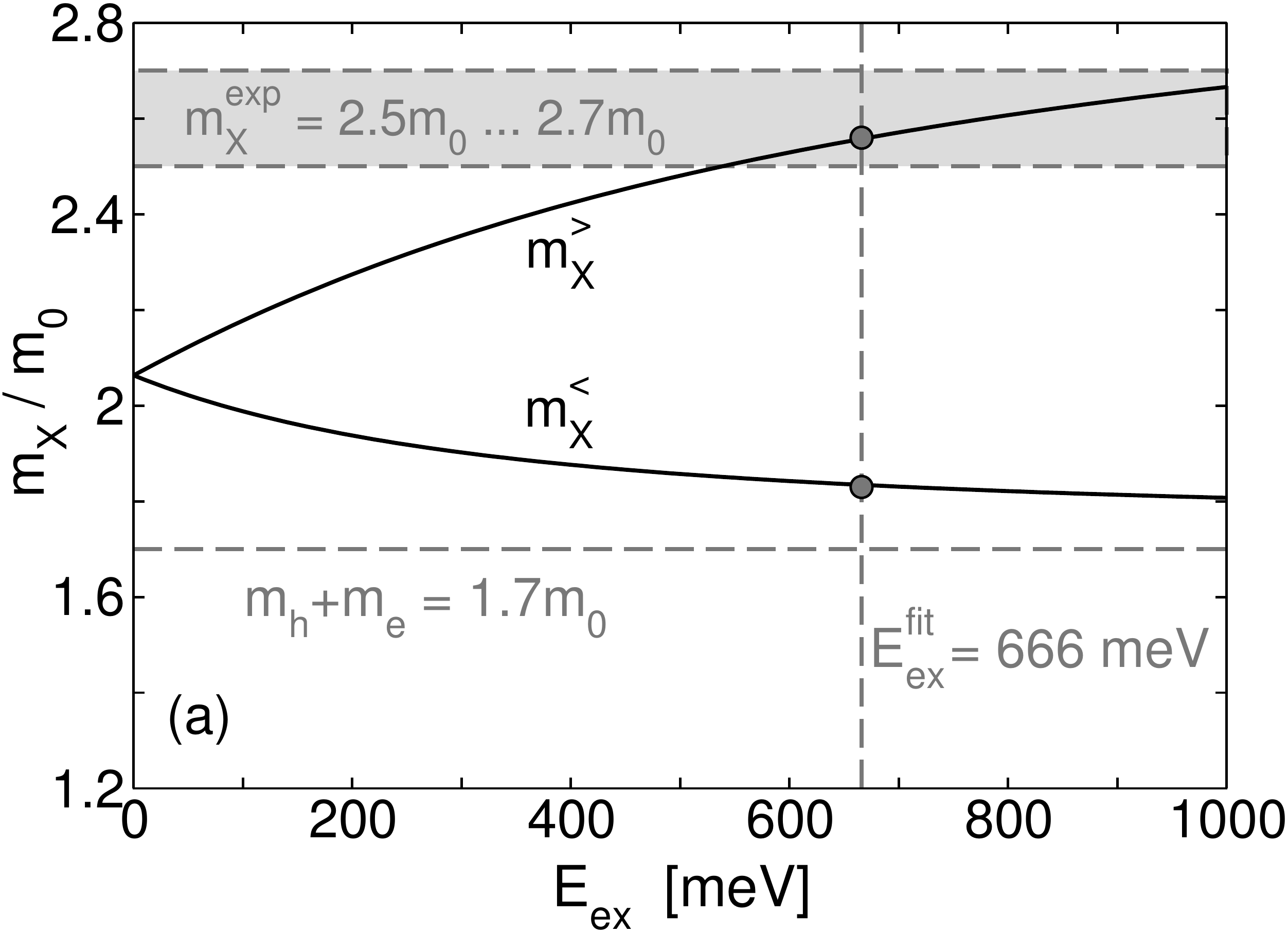}
\hspace*{\fill}
\raisebox{3.5pt}{\includegraphics[scale=0.23]{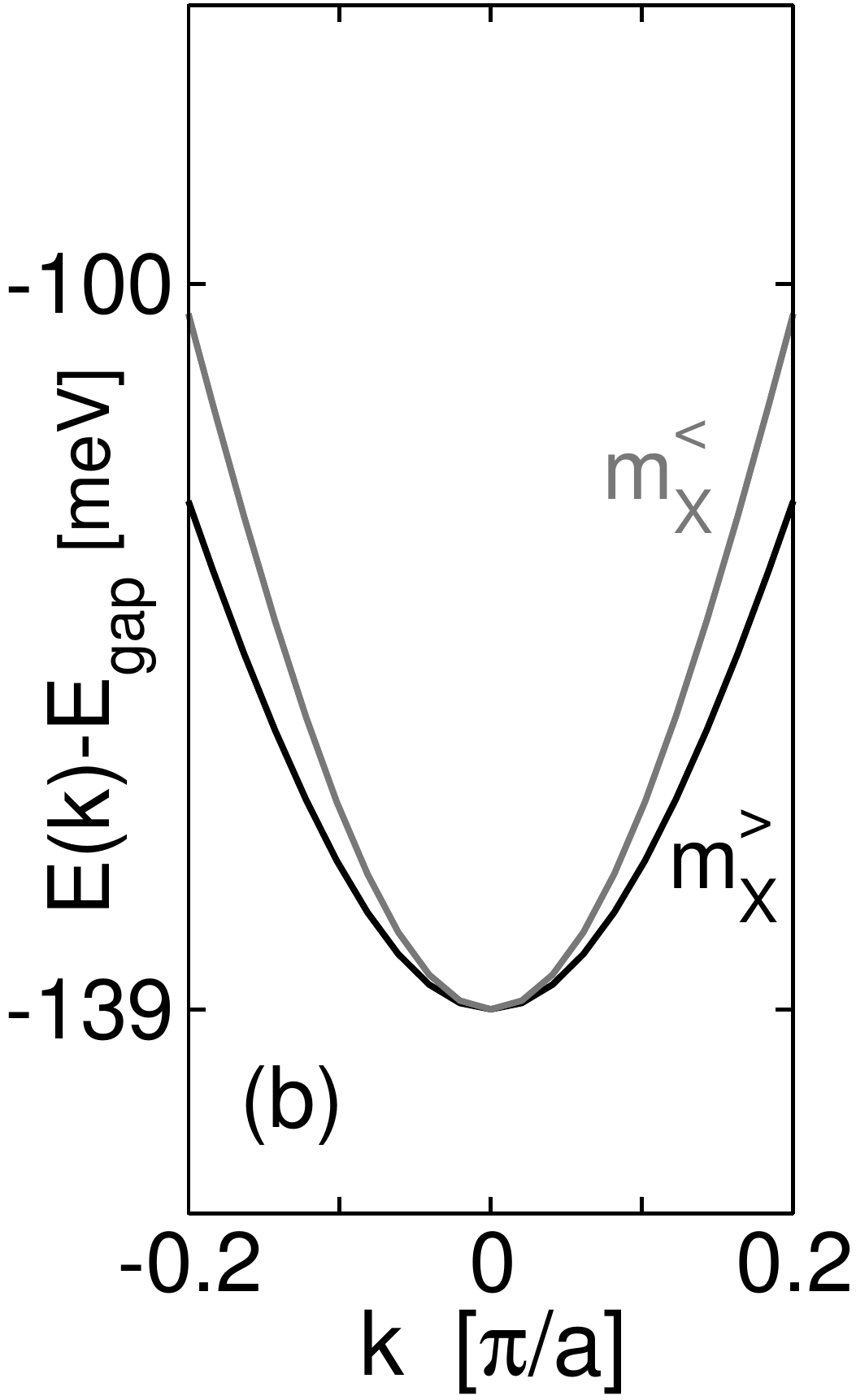}}
\hspace*{\fill}
\caption{Left panel (a): Exciton mass $m_X^<$, $m_X^>$ in the 1S ortho-exciton state, as a function of the exchange energy $E_\mathrm{ex}$. 
The value $m_X^< = m_X^>$ in the limit $E_\mathrm{ex}=0$ corresponds to the exciton mass in the 1S para-exciton state.
The vertical dashed line is placed at the value $E^\mathrm{fit}_\mathrm{ex} = 666\,\mathrm{meV}$ obtained from the previous fit to the spectrum, which results in $m_X^< = 1.83 m_0$, $m_X^> = 2.56 m_0$.
The horizontal dashed lines give the value of $m_X = m_h + m_e$ that would be obtained within Wannier theory, and the range of experimental values $2.5 \lesssim m_X / m_0 \lesssim 2.7$.
Right panel (b): Exciton dispersion $E(k)$ as a function of momentum $k$ parallel to a lattice axis, for the 1S ortho-exciton state.}
\label{fig:MassCuprous}
\end{figure}

\subsection{Exciton mass and spectrum}

With the model parameters from Tab.~\ref{tab:CUO_params} we can now compute, without further adjustments, the exciton mass and the remaining states in the exciton spectrum.
A few numerical values are listed in Tab.~\ref{tab:CUO_results}.

Fig.~\ref{fig:MassCuprous} shows the mass in the 1S ortho-exciton state.
The ortho-exciton state transforms as $\Gamma^+_5$ at $\vec K = 0$,
which splits into a one-dimensional and a two-dimensional representation at finite $\vec K$ parallel to a lattice axis (on a $\Delta$-line). 
Therefore, the exciton mass can assume two different values $m_X^< , m_X^>$.
They coincide only in the limit $E_\mathrm{ex}=0$, where they give the mass of the 1S para-exciton state ($2.06 \, m_0$),
and evolve in opposite directions as $E_\mathrm{ex}$ increases.
At the value $E_\mathrm{ex}=666\,\mathrm{meV}$ from the previous model parameter fit,
we obtain $m_X^> = 2.56 \, m_0$, which is in reasonable agreement with the experimental data (cf. Tab.~\ref{tab:CUO_results}).
Note the significant influence of electron-hole exchange on the exciton mass that can be observed in Fig.~\ref{fig:MassCuprous}: Without exchange, the correct exciton mass could not be obtained.

The right panel (b) in Fig.~\ref{fig:MassCuprous} shows the dispersion of the heavy ($m_X^>$) and light ($m_X^<$) exciton state.
We include this panel mainly to demonstrate that the exciton dispersion is parabolic at small $\vec K$, in contrast to the dispersion of a hole in the split-off valence band out of which this exciton state is formed.

\begin{figure}
\hspace*{\fill}
\includegraphics[scale=0.25]{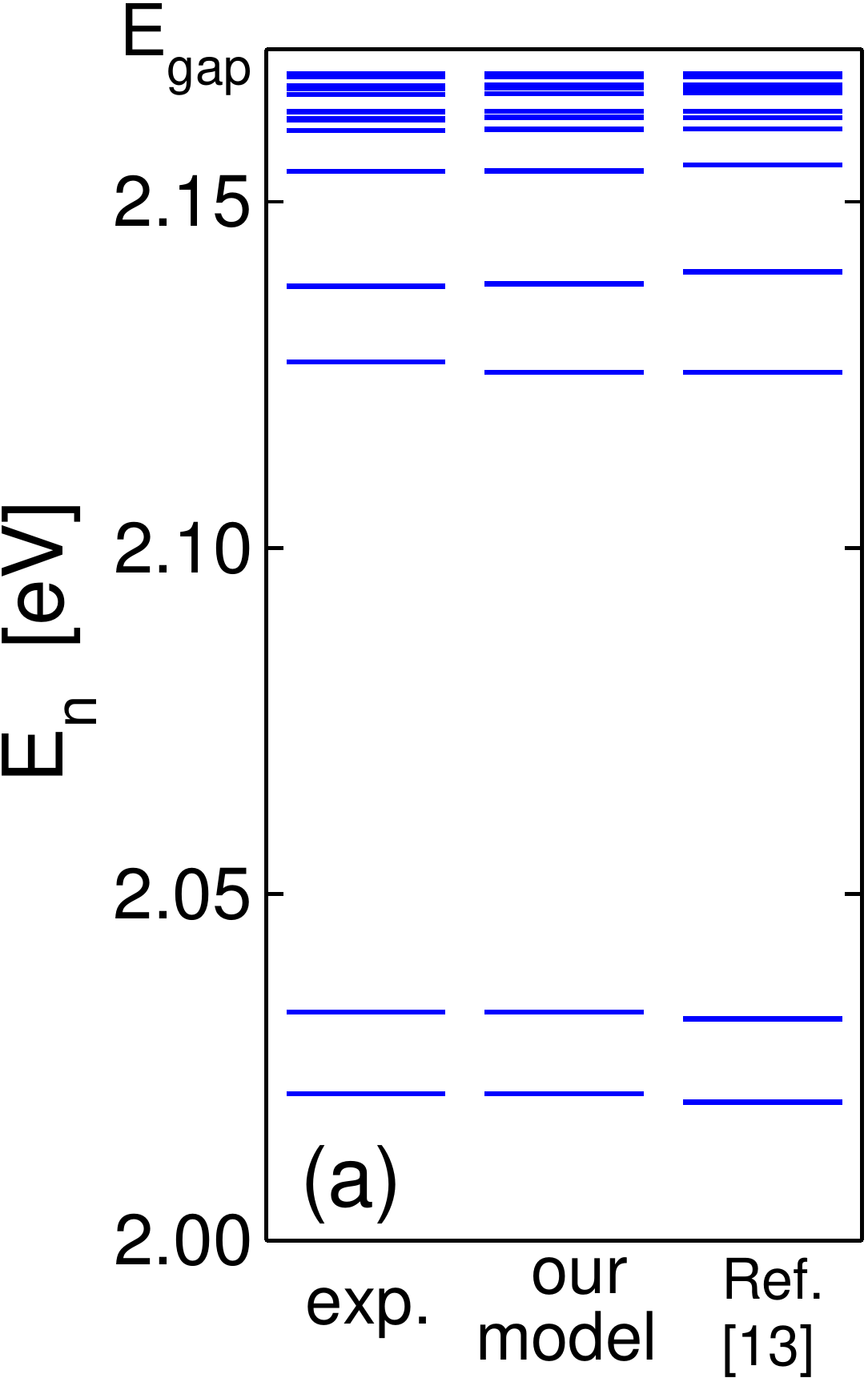}
\hspace*{\fill}
\includegraphics[scale=0.25]{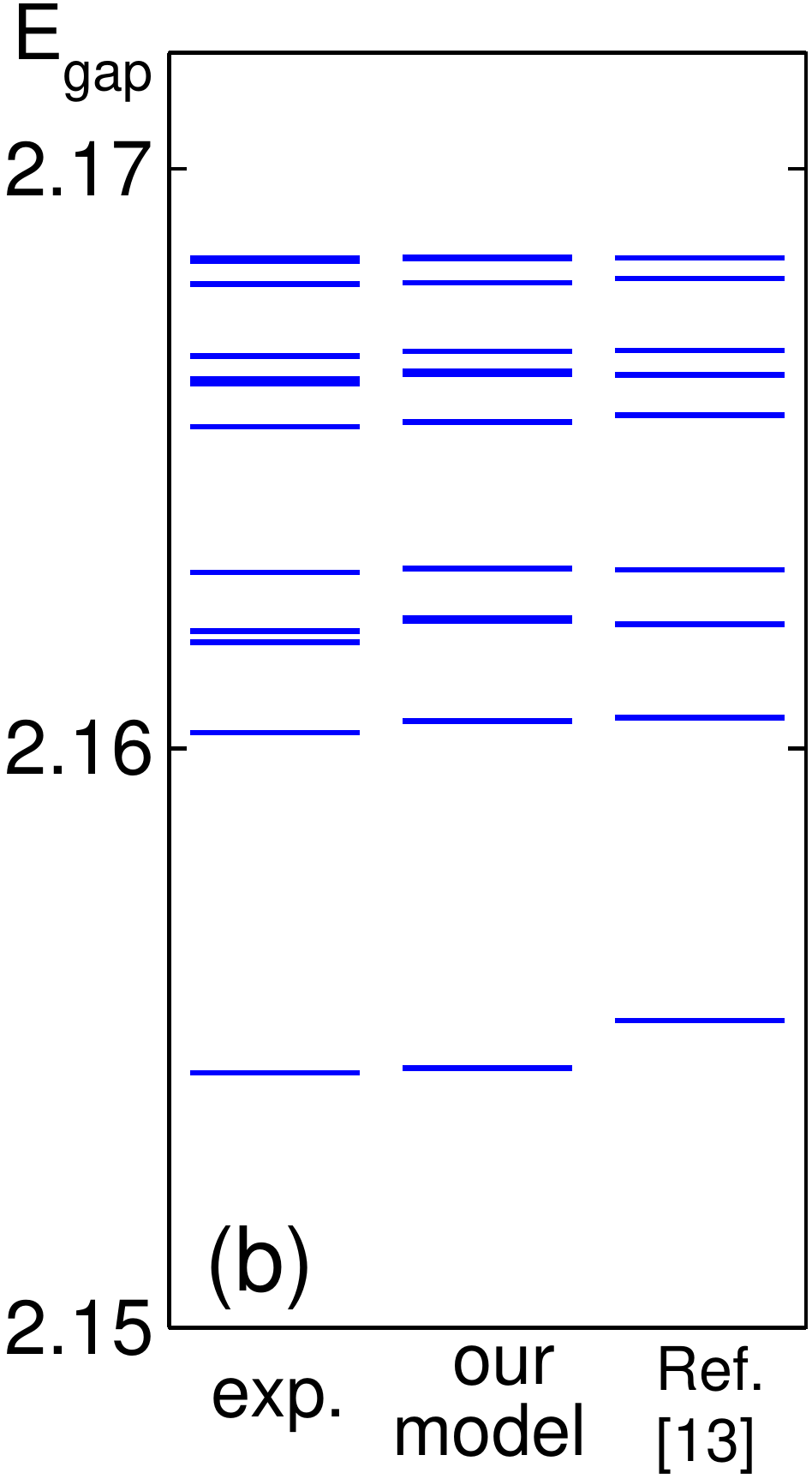}
\hspace*{\fill}
\includegraphics[scale=0.25]{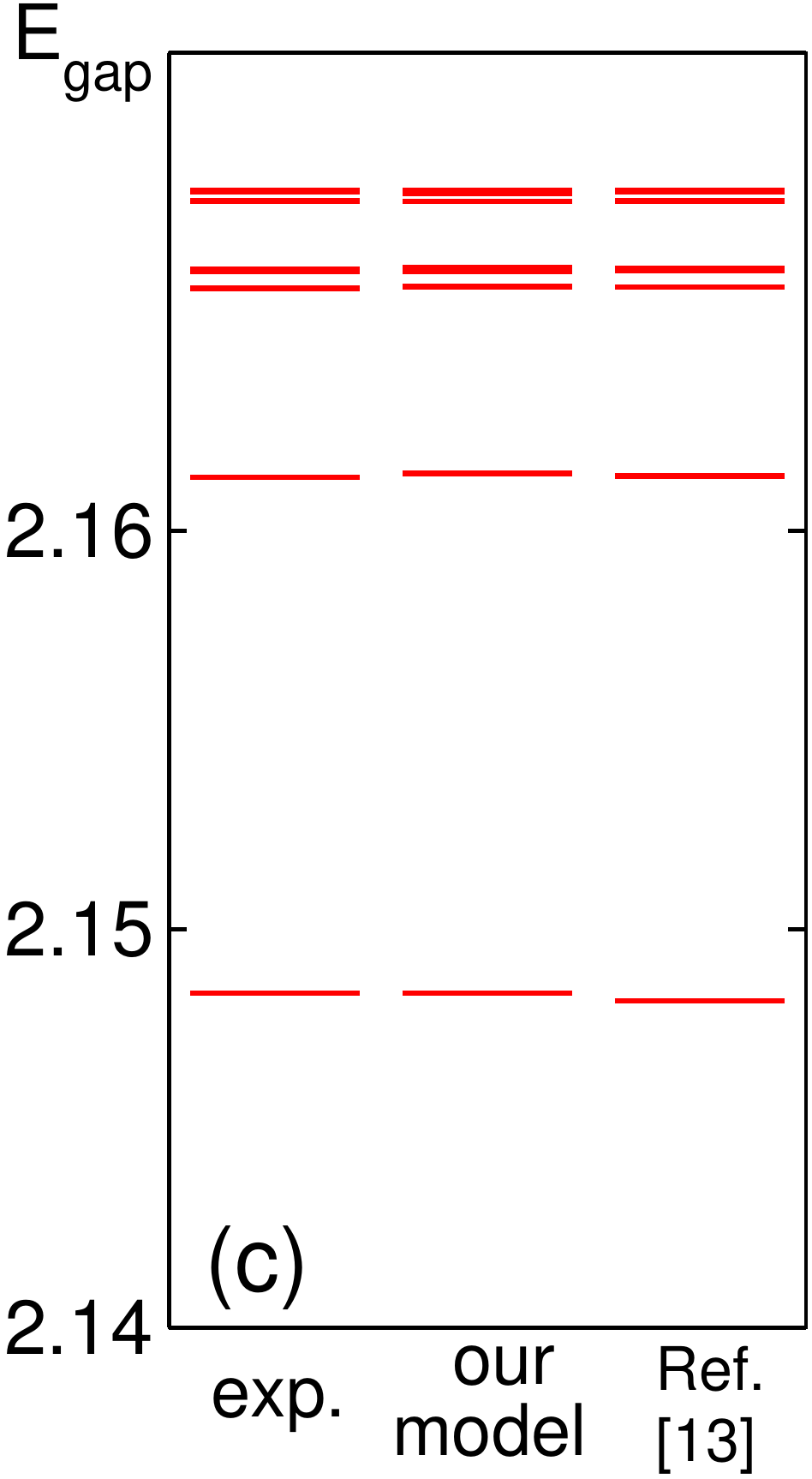}
\hspace*{\fill}
\caption{Exciton spectrum in the cuprous oxide, for even (left and central panels (a), (b)) and odd (rightmost panel (c)) exciton states up to $n \le 5$.
Each panel gives the experimental values (as in Fig.~\ref{fig:Cuprous}) in comparison to our model computation and the theoretical values from Ref.~\cite{SMWU17}.
Note the different scales of the energy axes. The central panel (b) is the magnification of the upper part of the left panel (a). 
}
\label{fig:SpecCuprous}
\end{figure}

Fig.~\ref{fig:SpecCuprous} shows the exciton spectrum. We compare the experimental data with our model computation and the spectrum from the continuum theory presented in Ref.~\cite{SMWU17}.
Agreement with experiment is comparably good for both the lattice model and the continuum theory, although not perfect in either case.
The maximal deviation is below $2 \, \mathrm{meV}$,
which is about $1.5 \%$ of the excitonic Rydberg energy. 
Both the continuum theory and the lattice model correctly reproduce the shift of exciton lines relative to the Rydberg series, and the splitting of exciton states in the cubic symmetry.

Note how these results prove the relevance of the exciton lattice model: The only information about the exciton spectrum put into the model is derived from the energy of the lowest three (two even and one odd) exciton states.
With this information, the model allows us to compute the entire spectrum with all deviations from the Rydberg series, and, as a second independent quantity, the exciton mass.
The agreement between model computations and experiment is certainly good enough to warrant the conclusion that all relevant aspects of the physics of exciton with strong-central cell corrections are correctly captured by the lattice model.

\begin{table}
\begin{tabular}{rrr}
\toprule
\noalign{\vskip1ex}
 &  experiment & our model 
 \\\noalign{\vskip1ex} \cmidrule(lr){1-3}\noalign{\vskip1ex}
 \multicolumn{1}{r}{1S ortho exc.} \\
mass $m_X/m_0$  &  \multicolumn{1}{c}{2.5 -- 2.7} & \multicolumn{1}{c}{$2.56$ ($m_X^>$)} \\[3pt]
 & & \multicolumn{1}{c}{$1.83$ ($m_X^<$)} \\[3pt]
radius & \multicolumn{1}{c}{---} & \multicolumn{1}{c}{$1.62 a \approx 7 \text{\AA}$}  \\[1.5ex]
\multicolumn{1}{r}{even states} \\
binding energy  & $151\phantom{.13}$ & --fitted-- \\
$E_\mathrm{gap} - E_n$ $[\mathrm{meV}]$ & $139\phantom{.13}$ & --fitted-- \\
 & $45.1\phantom{3}$ &  $46.62$ \\ 
 & $34.2\phantom{3}$ & $33.84$ \\
 & $17.6\phantom{3}$ & $17.52$ \\
 & $11.73$ & $11.53$ \\
& $10.17$ & $9.80$ \\
& $9.98$ & $9.75$ \\
& $8.97$ & $8.90$ \\
\noalign{\vskip1ex}
\bottomrule
\end{tabular}
\caption{Model results for excitons in the cuprous oxide $\mathrm{Cu}_2\mathrm{O}$,
including binding energies of the lowest even exciton states,
in comparison to experimental values.
}
\label{tab:CUO_results}
\end{table}

\section{Conclusion}
\label{sec:Conc}

Starting from a microscopic lattice model for exciton formation we study the effects of central-cell corrections on small-radius excitons, using the yellow exciton series in the cuprous oxide as the main example.
Our study relates two, at first sight unrelated, quantities: The shift
of even exciton states relative to the Rydberg series and the enhancement of the exciton mass. 

The success of the lattice model partly stems from the fact that it incorporates central-cell corrections in a natural way, such that specific exciton properties follow without additional effort.
Local corrections to the Coulomb attraction and non-parabolic terms in the kinetic energy are an intrinsic part of the model.
A crucial ingredient is spin-orbit coupling, which immediately explains the 
peculiar dispersion of the highest valence band.

Not surprisingly, exchange interaction is equally crucial, as it is responsible for the splitting of para-exciton and ortho-exciton states.
It is probably surprising, however, that exchange interaction is also responsible (i) for pushing the ortho-exciton mass towards the correct experimental value,
and (ii) for a strongly anisotropic ortho-exciton mass.
In particular, the mass enhancement for ortho-exciton and para-exciton is not the same.

The anisotropy of the ortho-exciton mass in the lattice model does not result from a $\vec k$-dependent~\cite{DFSKSB03,DFKSBS04,DFSKSB05} exchange interaction---exchange is local in our model---but from the interplay of exchange interaction with the other terms in the Hamiltonian, including spin-orbit coupling.
Furthermore, the mass anisotropy is not small: The `heavy' and `light' ortho-exciton mass differs by a factor $1.4$.
One consequence is that ortho-exciton states are no longer three-fold degenerate at finite $\vec k$.

In this context, note that the para-exciton mass obtained from the lattice model ($2.06 \, m_0$) is distinctly smaller than recent experimental values ($2.61 \, m_0$ in Ref.~\cite{BFSBSN07}), although it somewhat agrees with smaller values cited elsewhere ($2.2 \, m_0$ in Ref.~\cite{JFKB05}).
Given the fact that our result for the ortho-exciton mass agrees nicely with the experiment,
one should try to identify the origin of this discrepancy both theoretically and experimentally.

The lattice model in the form used here is kept deliberate simple, yet is easily extendable.
This poses a natural question:
Is it possible to perfect the agreement between theory and experiment? The answer is, of course, yes.
Reasonable extensions of the model would include additional terms in the hole kinetic energy (most importantly the terms with $\vec e=(0,1,1)$ and $\vec e = (1,1,1)$ in the notation of App.~\ref{app:Kinetic}), a refined treatment of screening of the Coulomb attraction, and Coulomb terms that couple different orbitals.

Inclusion of additional terms in the Hamiltonian introduces additional parameters.
Kinetic energy terms could be fitted to DFT band structure calculations or deduced from comparison to experimental exciton data.
The value $U(0)$ of the on-site Coulomb interaction could be estimated from the DFT orbitals,
and screening could be included via model potentials.
The present agreement between model and experiment is yet good enough to let us prefer the present simpler model over excessive parameter fitting.

As the principal usefulness of an exciton lattice model has now been established, several questions remain to be resolved.
For example, we have not computed the oscillator strengths of one- or two-photon absorption,
or the splitting of exciton states under mechanical stress~\cite{Tre77,WPBC80}.
The computation of oscillator strengths requires some additional theoretical work,
while the latter issue is immediately treatable by letting the kinetic energy terms become direction dependent.
Closer examination of these and related questions should be a worthwhile activity in future investigations.

\acknowledgments

This work was financed by Deutsche Forschungsgemeinschaft via AL1317/1-2 and SFB 652.
A.A. is grateful for helpful discussions with H. Knedlik.

\appendix

\section{Higher order terms of the kinetic energy}
\label{app:Kinetic}

The hole kinetic energy terms have the form
\begin{equation}
H_\mathrm{kin}[\vec e] =\frac{1}{|G_{\vec e} |} \sum_{g \in O_h} \Gamma(g) \, \mathsf H_{\vec e} \,\Gamma(g)^{-1} \, T(g \cdot \vec e) \;,
\end{equation}
with a translation vector $\vec e \in \mathbb Z^3$
and a $3 \times 3$ matrix $\mathsf H_{\vec e}$. 
The matrices $\Gamma(g)$ are given by the representation $\Gamma_5^+$ for the valence band orbitals,
and $g \cdot \vec e$ denotes the action of the group element $g \in O_h$ on a vector $\vec e$.
The prefactor $1/|G_{\vec e}|$, with the isotropy group $G_{\vec e} = \{ g \in O_h \text{ with } g \cdot \vec e = \vec e \}$, corrects for multiple counting.

Hermiticity of $H_\mathrm{kin}[\vec e]$ and the existence of an inversion $g_\mathrm{i} \in O_h$ with $g_\mathrm{i} \cdot \vec e = - \vec e$ implies $\mathsf H_{\vec e} = \Gamma(g_\mathrm{i}) \, \mathsf H_{\vec e} \Gamma(g_\mathrm{i})^{-1}$. For the $\Gamma_5^+$ representation, $\Gamma(g_\mathrm{i}) = 1$,
such that the matrix $\mathsf H_{\vec e}$ itself must be Hermitian.

Furthermore, we must have $\Gamma(g) \mathsf H_{\vec e} \Gamma(g)^{-1} = \mathsf H_{\vec e}$ for every $g \in G_{\vec e}$.
Constraints on $\mathsf H_{\vec e}$ result from elements $g \ne 1$ of $G_{\vec e}$.
To enumerate the different cases, note that among the vectors $g \cdot \vec e$ we can always find one vector whose entries are in ascending order $e_x \le e_y \le e_z$.
Apart from the trivial case $\vec e = 0$, we have to consider the six cases listed in Tab.~\ref{tab:Symm}.
Note that we do not list the case $0 < e_x = e_y < e_z$,
since it is equivalent to the fifth case ($0 < e_x < e_y = e_z$) up to relabeling of the coordinates.

\begin{table}
\hspace*{\fill}
\begin{tabular}{ccc}
\toprule
\noalign{\vskip1ex}
$\vec e$ & $|G_{\vec e}|$ & $\mathsf H_{\vec e}$ \\
\noalign{\vskip1ex}
\cmidrule(lr){1-3}
\noalign{\vskip1ex}
\rule{20ex}{0pt} & \rule{15ex}{0pt} & \rule{15ex}{0pt} \\[-1ex]
$0 = e_x = e_y < e_z$ & 8 & \large$\left(\begin{smallmatrix} a\, &  &  \\[1pt]  & a\, &  \\[1pt]  &  & b\,  \end{smallmatrix}\right)$ \\[3ex]
$0 = e_x < e_y = e_z$ & 4 & \large$\left(\begin{smallmatrix} a\, &  &  \\[1pt]  & b\, & c\,  \\[1pt]  &  c\, & b\,  \end{smallmatrix}\right)$ \\[3ex]
$0 = e_x < e_y < e_z$ & 2 & \large$\left(\begin{smallmatrix} a\, &  &  \\[1pt]  & b\, & d^*\!\!\!\,  \\[1pt]  &  d\, & c\,  \end{smallmatrix}\right)$ \\[3ex]
$0 < e_x = e_y = e_z$ & 6 & \large$\left(\begin{smallmatrix} a\, & b\, & \bar b\,  \\[1pt]  b\, & a\, & b\,  \\[1pt] \bar b\,  &  b\,  & a\,  \end{smallmatrix}\right)$ \\[3ex]
$0 < e_x < e_y = e_z$ & 2 & \large$\left(\begin{smallmatrix} a \, & d^*\!\!\!\, & \bar d^*\!\!\!\,  \\[1pt]  d\, & b \,  & c\,  \\[1pt] \bar d \,  &  c\, & b \,  \end{smallmatrix}\right)$ \\[3ex]
$0 < e_x < e_y < e_z$ & 1 & \large$\left(\begin{smallmatrix} a\, & d^*\!\!\!\, & e^*\!\!\!\, \\[1.5pt] d\, & b & f^*\!\!\!\, \\[1pt] e\, & f\, & c  \end{smallmatrix}\right)$
\\
\noalign{\vskip2ex}\bottomrule
\end{tabular} 
\hspace*{\fill}
\caption{
Possible hole kinetic energy terms for the $\Gamma_5^+$ representation.
In the third column, equal elements of $\mathsf H_e$ are denoted by the same letter, with $a,b,c \in \mathbb R$ and $d,e,f \in \mathbb C$,
a bar indicates negation as in $\bar a = -a$, and an empty entry indicates a zero element.
}
\label{tab:Symm}
\end{table}

\section{Hole dispersion}
\label{app:HoleBands}

The hole Hamiltonian $H_\mathrm{hole} + H_\mathrm{so}$ commutes
with each operator
$R^h_\mathsf{d} = (2 I_\mathsf{d}^2 - 1) \otimes (2/\hbar) S^h_\mathsf{d}$,
for $\mathsf d = x,y,z$,
such that its eigenstates can be classified according to one of these operators, say, for $\mathsf d = z$.
Note that, in contrast to the operators $R_\mathsf{d}$ from Sec.~\ref{sec:X},
the operators $R^h_\mathsf{d}$ do not commute among themselves.
The corresponding Bloch Hamiltonian, parametrized by the hole momentum $\vec k_h$,
splits in two $3 \times 3$ blocks:
One in the subspace of states $|\breve x, \uparrow\rangle$, $|\breve y, \uparrow\rangle$, $|\breve z, \downarrow\rangle$ to the eigenvalue $+1$ of $R^h_z$, the other in the subspace of states
$|\breve x, \downarrow\rangle$, $|\breve y, \downarrow\rangle$, $|\breve z, \uparrow\rangle$
to the eigenvalue $-1$ of $R^h_z$.
Both blocks are related by the unitary transformation induced by $R^h_x$ (or $R^h_y$),
and have the same eigenvalues.
Therefore, the hole bands are two-fold degenerate.

For $k_y = k_z = 0$, diagonalization of one of the $3 \times 3$ matrices gives the eigenvalues
\begin{equation}
\begin{split}
 E_1(k_x) &= E_v+ 2 t_2 (c_x - 1) + 2 \lambda \;, \\
 E_2(k_x) &= E_v +  (t_1 + t_2) (c_x - 1) - \lambda - \xi \;, \\
  E_3(k_x) &= E_v + (t_1 + t_2) (c_x - 1) - \lambda + \xi \;, 
\end{split}
\end{equation}
with $c_\mathsf{d} = \cos a (\vec k_h \cdot \vec e_\mathsf{d})$, $\lambda = - \frac{E_\mathrm{so}}{6}$, $E_v = 2 t_1 + 4 t_2$, and $\xi^2 =   \left( (t_1 - t_2)(c_x - 1) + \lambda \right)^2 + 8 \lambda^2 $.
Equivalent expressions hold for the $y$, $z$ direction.
At $\vec k=0$, with $E_1(0) = E_2(0) = E_v - (1/3) E_\mathrm{so}$,
$E_3(0) = E_v + (2/3) E_\mathrm{so}$, the energy level splitting is $E_\mathrm{so}$.

\begin{table}[b]
\begin{tabular}{ccc} 
\toprule\noalign{\vskip2ex}
\rule{1ex}{0pt}
$R_x \; R_y \; R_z$ & \rule{2ex}{0pt} states $(\times \sqrt 2)$ \rule{2ex}{0pt}  & \rule{2ex}{0pt} $\left( \dfrac14 - \dfrac{1}{\hbar^2} \vec S^h \cdot \vec S^e \right)_\mathrm{diag}$ \rule{2ex}{0pt} 
 \\\noalign{\vskip2ex}
 \cmidrule(lr){1-3}
 \noalign{\vskip2ex}
$ + \; + \; + $
 & $\begin{matrix}|\breve x {\uparrow\uparrow}\rangle - |\breve x{\downarrow\downarrow}\rangle \\ |\breve y{\uparrow\uparrow}\rangle + |\breve y{\downarrow\downarrow}\rangle \\   |\breve z{\downarrow\uparrow}\rangle + |\breve z{\uparrow\downarrow}\rangle \end{matrix}$ & 
$\begin{matrix} 0 \\ 0 \\ 0 \end{matrix}$
  \\[5ex]
$ - \; - \; + $
 &
$\begin{matrix}|\breve x{\uparrow\uparrow}\rangle + |\breve x{\downarrow\downarrow}\rangle \\ |\breve y{\uparrow\uparrow}\rangle - |\breve y{\downarrow\downarrow}\rangle \\   |\breve z{\downarrow\uparrow}\rangle -|\breve z{\uparrow\downarrow}\rangle  \end{matrix}$ & 
$\begin{matrix} 0 \\ 0 \\ 1 \end{matrix}$
\\[5ex]
$ - \; + \; - $
 &
$\begin{matrix}|\breve x{\uparrow\downarrow}\rangle  + |\breve x{\downarrow\uparrow}\rangle \\ |\breve y{\uparrow\downarrow}\rangle - |\breve y{\downarrow\uparrow}\rangle \\   |\breve z{\downarrow\downarrow}\rangle  - |\breve z{\uparrow\uparrow}\rangle  
\end{matrix}$ & 
$\begin{matrix} 0  \\ 1 \\ 0 \end{matrix}$
\\[5ex]
$ + \; - \; - $ &
$\begin{matrix}|\breve x{\uparrow\downarrow}\rangle - |\breve x{\downarrow\uparrow}\rangle \\ |\breve y{\uparrow\downarrow}\rangle + |\breve y{\downarrow\uparrow}\rangle \\   |\breve z{\downarrow\downarrow}\rangle + |\breve z{\uparrow\uparrow}\rangle \end{matrix}$ & 
$\begin{matrix} 1 \\ 0  \\ 0 \end{matrix}$
\\\noalign{\vskip2ex}\bottomrule
\end{tabular}
\caption{Symmetrized basis states with respect to the operators $R_\mathsf{d}$ from Sec.~\ref{sec:X}.}
\label{tab:SymStates}
\end{table}

\section{Symmetrized basis states}
\label{app:basis}

The local degrees of freedom of the Hamiltonian $H_X$ in Eq.~\eqref{HX}
are labelled by the orbital, hole spin, and electron spin indices as
$|\breve{\mathsf d}, \mathsf s_h, \mathsf s_e \rangle$ with $\mathsf d = x,y,z$ and $\mathsf s_h, \mathsf s_e = \uparrow, \downarrow$ (cf. Sec.~\ref{sec:LKM}).
Symmetrized basis states with respect to the operators $R_\mathsf{d}$ from Sec.~\ref{sec:X} are given in Tab.~\ref{tab:SymStates}. They belong to four different sets depending on the eigenvalues $\pm 1$ of the $R_\mathsf{d}$ operators. Since $R_x  R_y R_z = 1$ only the listed eigenvalue combinations can occur.
The first set gives the para-exciton states,
and contains only basis states with non-zero hole-electron spin $\vec S^h + \vec S^e$,
 the three remaining sets give the ortho-exciton states.

Spin-orbit and exchange coupling connect only the three states within each group.
For each group,
spin-orbit coupling is given by the same $3 \times 3$ matrix
\begin{equation}
 \frac{2}{\hbar} \vec I \cdot \vec S^h \equiv \begin{pmatrix} \phantom- 0 & \phantom- \ii & \phantom- 1 \\ - \ii & \phantom- 0 & \phantom- \ii \\ \phantom- 1 & -\ii & \phantom- 0 \end{pmatrix} \;,
\end{equation}
while exchange interaction is given by a diagonal matrix with entries as listed in the last column of Tab.~\ref{tab:SymStates}.


\begin{thebibliography}{34}%
\makeatletter
\providecommand \@ifxundefined [1]{%
 \@ifx{#1\undefined}
}%
\providecommand \@ifnum [1]{%
 \ifnum #1\expandafter \@firstoftwo
 \else \expandafter \@secondoftwo
 \fi
}%
\providecommand \@ifx [1]{%
 \ifx #1\expandafter \@firstoftwo
 \else \expandafter \@secondoftwo
 \fi
}%
\providecommand \natexlab [1]{#1}%
\providecommand \enquote  [1]{``#1''}%
\providecommand \bibnamefont  [1]{#1}%
\providecommand \bibfnamefont [1]{#1}%
\providecommand \citenamefont [1]{#1}%
\providecommand \href@noop [0]{\@secondoftwo}%
\providecommand \href [0]{\begingroup \@sanitize@url \@href}%
\providecommand \@href[1]{\@@startlink{#1}\@@href}%
\providecommand \@@href[1]{\endgroup#1\@@endlink}%
\providecommand \@sanitize@url [0]{\catcode `\\12\catcode `\$12\catcode
  `\&12\catcode `\#12\catcode `\^12\catcode `\_12\catcode `\%12\relax}%
\providecommand \@@startlink[1]{}%
\providecommand \@@endlink[0]{}%
\providecommand \url  [0]{\begingroup\@sanitize@url \@url }%
\providecommand \@url [1]{\endgroup\@href {#1}{\urlprefix }}%
\providecommand \urlprefix  [0]{URL }%
\providecommand \Eprint [0]{\href }%
\providecommand \doibase [0]{http://dx.doi.org/}%
\providecommand \selectlanguage [0]{\@gobble}%
\providecommand \bibinfo  [0]{\@secondoftwo}%
\providecommand \bibfield  [0]{\@secondoftwo}%
\providecommand \translation [1]{[#1]}%
\providecommand \BibitemOpen [0]{}%
\providecommand \bibitemStop [0]{}%
\providecommand \bibitemNoStop [0]{.\EOS\space}%
\providecommand \EOS [0]{\spacefactor3000\relax}%
\providecommand \BibitemShut  [1]{\csname bibitem#1\endcsname}%
\let\auto@bib@innerbib\@empty
\bibitem [{\citenamefont {Knox}(1963)}]{Knox63}%
  \BibitemOpen
  \bibfield  {author} {\bibinfo {author} {\bibfnamefont {R.}~\bibnamefont
  {Knox}},\ }\href@noop {} {\emph {\bibinfo {title} {Theory of excitons}}}\
  (\bibinfo  {publisher} {Academic, New York},\ \bibinfo {year}
  {1963})\BibitemShut {NoStop}%
\bibitem [{\citenamefont {Moskalenko}\ and\ \citenamefont
  {Snoke}(2000)}]{MS00}%
  \BibitemOpen
  \bibfield  {author} {\bibinfo {author} {\bibfnamefont {S.~A.}\ \bibnamefont
  {Moskalenko}}\ and\ \bibinfo {author} {\bibfnamefont {D.~W.}\ \bibnamefont
  {Snoke}},\ }\href@noop {} {\emph {\bibinfo {title} {{Bose}-{Einstein}
  Condensation of Excitons and Biexcitons}}}\ (\bibinfo  {publisher} {Cambridge
  University Press},\ \bibinfo {year} {2000})\BibitemShut {NoStop}%
\bibitem [{\citenamefont {Kazimierczuk}\ \emph {et~al.}(2014)\citenamefont
  {Kazimierczuk}, \citenamefont {Fr{\"o}hlich}, \citenamefont {Scheel},
  \citenamefont {Stolz},\ and\ \citenamefont {Bayer}}]{KFSSB14}%
  \BibitemOpen
  \bibfield  {author} {\bibinfo {author} {\bibfnamefont {T.}~\bibnamefont
  {Kazimierczuk}}, \bibinfo {author} {\bibfnamefont {D.}~\bibnamefont
  {Fr{\"o}hlich}}, \bibinfo {author} {\bibfnamefont {S.}~\bibnamefont
  {Scheel}}, \bibinfo {author} {\bibfnamefont {H.}~\bibnamefont {Stolz}}, \
  and\ \bibinfo {author} {\bibfnamefont {M.}~\bibnamefont {Bayer}},\
  }\href@noop {} {\bibfield  {journal} {\bibinfo  {journal} {Nature}\ }\textbf
  {\bibinfo {volume} {514}},\ \bibinfo {pages} {343} (\bibinfo {year}
  {2014})}\BibitemShut {NoStop}%
\bibitem [{\citenamefont {Kavoulakis}\ \emph {et~al.}(1997)\citenamefont
  {Kavoulakis}, \citenamefont {Chang},\ and\ \citenamefont {Baym}}]{KCB97}%
  \BibitemOpen
  \bibfield  {author} {\bibinfo {author} {\bibfnamefont {G.~M.}\ \bibnamefont
  {Kavoulakis}}, \bibinfo {author} {\bibfnamefont {Y.-C.}\ \bibnamefont
  {Chang}}, \ and\ \bibinfo {author} {\bibfnamefont {G.}~\bibnamefont {Baym}},\
  }\href@noop {} {\bibfield  {journal} {\bibinfo  {journal} {Phys. Rev. B}\
  }\textbf {\bibinfo {volume} {55}},\ \bibinfo {pages} {7593} (\bibinfo {year}
  {1997})}\BibitemShut {NoStop}%
\bibitem [{\citenamefont {Fr{\"o}hlich}\ \emph {et~al.}(1979)\citenamefont
  {Fr{\"o}hlich}, \citenamefont {Kenklies}, \citenamefont {Uihlein},\ and\
  \citenamefont {Schwab}}]{FKUS79}%
  \BibitemOpen
  \bibfield  {author} {\bibinfo {author} {\bibfnamefont {D.}~\bibnamefont
  {Fr{\"o}hlich}}, \bibinfo {author} {\bibfnamefont {R.}~\bibnamefont
  {Kenklies}}, \bibinfo {author} {\bibfnamefont {C.}~\bibnamefont {Uihlein}}, \
  and\ \bibinfo {author} {\bibfnamefont {C.}~\bibnamefont {Schwab}},\
  }\href@noop {} {\bibfield  {journal} {\bibinfo  {journal} {Phys. Rev. Lett.}\
  }\textbf {\bibinfo {volume} {43}},\ \bibinfo {pages} {1260} (\bibinfo {year}
  {1979})}\BibitemShut {NoStop}%
\bibitem [{\citenamefont {Uihlein}\ \emph {et~al.}(1981)\citenamefont
  {Uihlein}, \citenamefont {Fr{\"o}hlich},\ and\ \citenamefont
  {Kenklies}}]{UFK81}%
  \BibitemOpen
  \bibfield  {author} {\bibinfo {author} {\bibfnamefont {C.}~\bibnamefont
  {Uihlein}}, \bibinfo {author} {\bibfnamefont {D.}~\bibnamefont
  {Fr{\"o}hlich}}, \ and\ \bibinfo {author} {\bibfnamefont {R.}~\bibnamefont
  {Kenklies}},\ }\href@noop {} {\bibfield  {journal} {\bibinfo  {journal}
  {Phys. Rev. B}\ }\textbf {\bibinfo {volume} {23}},\ \bibinfo {pages} {2731}
  (\bibinfo {year} {1981})}\BibitemShut {NoStop}%
\bibitem [{\citenamefont {Thewes}\ \emph {et~al.}(2015)\citenamefont {Thewes},
  \citenamefont {Heck\"otter}, \citenamefont {Kazimierczuk}, \citenamefont
  {A\ss{}mann}, \citenamefont {Fr\"ohlich}, \citenamefont {Bayer},
  \citenamefont {Semina},\ and\ \citenamefont {Glazov}}]{THKAFBSG12}%
  \BibitemOpen
  \bibfield  {author} {\bibinfo {author} {\bibfnamefont {J.}~\bibnamefont
  {Thewes}}, \bibinfo {author} {\bibfnamefont {J.}~\bibnamefont {Heck\"otter}},
  \bibinfo {author} {\bibfnamefont {T.}~\bibnamefont {Kazimierczuk}}, \bibinfo
  {author} {\bibfnamefont {M.}~\bibnamefont {A\ss{}mann}}, \bibinfo {author}
  {\bibfnamefont {D.}~\bibnamefont {Fr\"ohlich}}, \bibinfo {author}
  {\bibfnamefont {M.}~\bibnamefont {Bayer}}, \bibinfo {author} {\bibfnamefont
  {M.~A.}\ \bibnamefont {Semina}}, \ and\ \bibinfo {author} {\bibfnamefont
  {M.~M.}\ \bibnamefont {Glazov}},\ }\href {\doibase
  10.1103/PhysRevLett.115.027402} {\bibfield  {journal} {\bibinfo  {journal}
  {Phys. Rev. Lett.}\ }\textbf {\bibinfo {volume} {115}},\ \bibinfo {pages}
  {027402} (\bibinfo {year} {2015})}\BibitemShut {NoStop}%
\bibitem [{\citenamefont {Schweiner}\ \emph {et~al.}(2016)\citenamefont
  {Schweiner}, \citenamefont {Main}, \citenamefont {Feldmaier}, \citenamefont
  {Wunner},\ and\ \citenamefont {Uihlein}}]{SMFWU16}%
  \BibitemOpen
  \bibfield  {author} {\bibinfo {author} {\bibfnamefont {F.}~\bibnamefont
  {Schweiner}}, \bibinfo {author} {\bibfnamefont {J.}~\bibnamefont {Main}},
  \bibinfo {author} {\bibfnamefont {M.}~\bibnamefont {Feldmaier}}, \bibinfo
  {author} {\bibfnamefont {G.}~\bibnamefont {Wunner}}, \ and\ \bibinfo {author}
  {\bibfnamefont {C.}~\bibnamefont {Uihlein}},\ }\href {\doibase
  10.1103/PhysRevB.93.195203} {\bibfield  {journal} {\bibinfo  {journal} {Phys.
  Rev. B}\ }\textbf {\bibinfo {volume} {93}},\ \bibinfo {pages} {195203}
  (\bibinfo {year} {2016})}\BibitemShut {NoStop}%
\bibitem [{\citenamefont {Sch\"one}\ \emph {et~al.}(2016)\citenamefont
  {Sch\"one}, \citenamefont {Kr\"uger}, \citenamefont {Gr\"unwald},
  \citenamefont {Stolz}, \citenamefont {Scheel}, \citenamefont {A\ss{}mann},
  \citenamefont {Heck\"otter}, \citenamefont {Thewes}, \citenamefont
  {Fr\"ohlich},\ and\ \citenamefont {Bayer}}]{SKGSSAHTFB16}%
  \BibitemOpen
  \bibfield  {author} {\bibinfo {author} {\bibfnamefont {F.}~\bibnamefont
  {Sch\"one}}, \bibinfo {author} {\bibfnamefont {S.-O.}\ \bibnamefont
  {Kr\"uger}}, \bibinfo {author} {\bibfnamefont {P.}~\bibnamefont
  {Gr\"unwald}}, \bibinfo {author} {\bibfnamefont {H.}~\bibnamefont {Stolz}},
  \bibinfo {author} {\bibfnamefont {S.}~\bibnamefont {Scheel}}, \bibinfo
  {author} {\bibfnamefont {M.}~\bibnamefont {A\ss{}mann}}, \bibinfo {author}
  {\bibfnamefont {J.}~\bibnamefont {Heck\"otter}}, \bibinfo {author}
  {\bibfnamefont {J.}~\bibnamefont {Thewes}}, \bibinfo {author} {\bibfnamefont
  {D.}~\bibnamefont {Fr\"ohlich}}, \ and\ \bibinfo {author} {\bibfnamefont
  {M.}~\bibnamefont {Bayer}},\ }\href {\doibase 10.1103/PhysRevB.93.075203}
  {\bibfield  {journal} {\bibinfo  {journal} {Phys. Rev. B}\ }\textbf {\bibinfo
  {volume} {93}},\ \bibinfo {pages} {075203} (\bibinfo {year}
  {2016})}\BibitemShut {NoStop}%
\bibitem [{\citenamefont {J\"orger}\ \emph {et~al.}(2005)\citenamefont
  {J\"orger}, \citenamefont {Fleck}, \citenamefont {Klingshirn},\ and\
  \citenamefont {von Baltz}}]{JFKB05}%
  \BibitemOpen
  \bibfield  {author} {\bibinfo {author} {\bibfnamefont {M.}~\bibnamefont
  {J\"orger}}, \bibinfo {author} {\bibfnamefont {T.}~\bibnamefont {Fleck}},
  \bibinfo {author} {\bibfnamefont {C.}~\bibnamefont {Klingshirn}}, \ and\
  \bibinfo {author} {\bibfnamefont {R.}~\bibnamefont {von Baltz}},\ }\href
  {\doibase 10.1103/PhysRevB.71.235210} {\bibfield  {journal} {\bibinfo
  {journal} {Phys. Rev. B}\ }\textbf {\bibinfo {volume} {71}},\ \bibinfo
  {pages} {235210} (\bibinfo {year} {2005})}\BibitemShut {NoStop}%
\bibitem [{\citenamefont {Brandt}\ \emph {et~al.}(2007)\citenamefont {Brandt},
  \citenamefont {Fr\"ohlich}, \citenamefont {Sandfort}, \citenamefont {Bayer},
  \citenamefont {Stolz},\ and\ \citenamefont {Naka}}]{BFSBSN07}%
  \BibitemOpen
  \bibfield  {author} {\bibinfo {author} {\bibfnamefont {J.}~\bibnamefont
  {Brandt}}, \bibinfo {author} {\bibfnamefont {D.}~\bibnamefont {Fr\"ohlich}},
  \bibinfo {author} {\bibfnamefont {C.}~\bibnamefont {Sandfort}}, \bibinfo
  {author} {\bibfnamefont {M.}~\bibnamefont {Bayer}}, \bibinfo {author}
  {\bibfnamefont {H.}~\bibnamefont {Stolz}}, \ and\ \bibinfo {author}
  {\bibfnamefont {N.}~\bibnamefont {Naka}},\ }\href {\doibase
  10.1103/PhysRevLett.99.217403} {\bibfield  {journal} {\bibinfo  {journal}
  {Phys. Rev. Lett.}\ }\textbf {\bibinfo {volume} {99}},\ \bibinfo {pages}
  {217403} (\bibinfo {year} {2007})}\BibitemShut {NoStop}%
\bibitem [{\citenamefont {Naka}\ \emph {et~al.}(2012)\citenamefont {Naka},
  \citenamefont {Akimoto}, \citenamefont {Shirai},\ and\ \citenamefont
  {Kan'no}}]{NASK12}%
  \BibitemOpen
  \bibfield  {author} {\bibinfo {author} {\bibfnamefont {N.}~\bibnamefont
  {Naka}}, \bibinfo {author} {\bibfnamefont {I.}~\bibnamefont {Akimoto}},
  \bibinfo {author} {\bibfnamefont {M.}~\bibnamefont {Shirai}}, \ and\ \bibinfo
  {author} {\bibfnamefont {K.-i.}\ \bibnamefont {Kan'no}},\ }\href {\doibase
  10.1103/PhysRevB.85.035209} {\bibfield  {journal} {\bibinfo  {journal} {Phys.
  Rev. B}\ }\textbf {\bibinfo {volume} {85}},\ \bibinfo {pages} {035209}
  (\bibinfo {year} {2012})}\BibitemShut {NoStop}%
\bibitem [{\citenamefont {Schweiner}\ \emph {et~al.}(2017)\citenamefont
  {Schweiner}, \citenamefont {Main}, \citenamefont {Wunner},\ and\
  \citenamefont {Uihlein}}]{SMWU17}%
  \BibitemOpen
  \bibfield  {author} {\bibinfo {author} {\bibfnamefont {F.}~\bibnamefont
  {Schweiner}}, \bibinfo {author} {\bibfnamefont {J.}~\bibnamefont {Main}},
  \bibinfo {author} {\bibfnamefont {G.}~\bibnamefont {Wunner}}, \ and\ \bibinfo
  {author} {\bibfnamefont {C.}~\bibnamefont {Uihlein}},\ }\href {\doibase
  10.1103/PhysRevB.95.195201} {\bibfield  {journal} {\bibinfo  {journal} {Phys.
  Rev. B}\ }\textbf {\bibinfo {volume} {95}},\ \bibinfo {pages} {195201}
  (\bibinfo {year} {2017})}\BibitemShut {NoStop}%
\bibitem [{\citenamefont {Luttinger}(1956)}]{Lutt56}%
  \BibitemOpen
  \bibfield  {author} {\bibinfo {author} {\bibfnamefont {J.~M.}\ \bibnamefont
  {Luttinger}},\ }\href {\doibase 10.1103/PhysRev.102.1030} {\bibfield
  {journal} {\bibinfo  {journal} {Phys. Rev.}\ }\textbf {\bibinfo {volume}
  {102}},\ \bibinfo {pages} {1030} (\bibinfo {year} {1956})}\BibitemShut
  {NoStop}%
\bibitem [{\citenamefont {Baldereschi}\ and\ \citenamefont
  {Lipari}(1971)}]{BL71}%
  \BibitemOpen
  \bibfield  {author} {\bibinfo {author} {\bibfnamefont {A.}~\bibnamefont
  {Baldereschi}}\ and\ \bibinfo {author} {\bibfnamefont {N.~C.}\ \bibnamefont
  {Lipari}},\ }\href {\doibase 10.1103/PhysRevB.3.439} {\bibfield  {journal}
  {\bibinfo  {journal} {Phys. Rev. B}\ }\textbf {\bibinfo {volume} {3}},\
  \bibinfo {pages} {439} (\bibinfo {year} {1971})}\BibitemShut {NoStop}%
\bibitem [{\citenamefont {Baldereschi}\ and\ \citenamefont
  {Lipari}(1973)}]{BL73}%
  \BibitemOpen
  \bibfield  {author} {\bibinfo {author} {\bibfnamefont {A.}~\bibnamefont
  {Baldereschi}}\ and\ \bibinfo {author} {\bibfnamefont {N.~O.}\ \bibnamefont
  {Lipari}},\ }\href {\doibase 10.1103/PhysRevB.8.2697} {\bibfield  {journal}
  {\bibinfo  {journal} {Phys. Rev. B}\ }\textbf {\bibinfo {volume} {8}},\
  \bibinfo {pages} {2697} (\bibinfo {year} {1973})}\BibitemShut {NoStop}%
\bibitem [{\citenamefont {Baldereschi}\ and\ \citenamefont
  {Lipari}(1974)}]{BL74}%
  \BibitemOpen
  \bibfield  {author} {\bibinfo {author} {\bibfnamefont {A.}~\bibnamefont
  {Baldereschi}}\ and\ \bibinfo {author} {\bibfnamefont {N.~O.}\ \bibnamefont
  {Lipari}},\ }\href {\doibase 10.1103/PhysRevB.9.1525} {\bibfield  {journal}
  {\bibinfo  {journal} {Phys. Rev. B}\ }\textbf {\bibinfo {volume} {9}},\
  \bibinfo {pages} {1525} (\bibinfo {year} {1974})}\BibitemShut {NoStop}%
\bibitem [{\citenamefont {Suzuki}\ and\ \citenamefont {Hensel}(1974)}]{SH74}%
  \BibitemOpen
  \bibfield  {author} {\bibinfo {author} {\bibfnamefont {K.}~\bibnamefont
  {Suzuki}}\ and\ \bibinfo {author} {\bibfnamefont {J.~C.}\ \bibnamefont
  {Hensel}},\ }\href {\doibase 10.1103/PhysRevB.9.4184} {\bibfield  {journal}
  {\bibinfo  {journal} {Phys. Rev. B}\ }\textbf {\bibinfo {volume} {9}},\
  \bibinfo {pages} {4184} (\bibinfo {year} {1974})}\BibitemShut {NoStop}%
\bibitem [{\citenamefont {Lipari}\ and\ \citenamefont
  {Altarelli}(1977)}]{LA77}%
  \BibitemOpen
  \bibfield  {author} {\bibinfo {author} {\bibfnamefont {N.~O.}\ \bibnamefont
  {Lipari}}\ and\ \bibinfo {author} {\bibfnamefont {M.}~\bibnamefont
  {Altarelli}},\ }\href {\doibase 10.1103/PhysRevB.15.4883} {\bibfield
  {journal} {\bibinfo  {journal} {Phys. Rev. B}\ }\textbf {\bibinfo {volume}
  {15}},\ \bibinfo {pages} {4883} (\bibinfo {year} {1977})}\BibitemShut
  {NoStop}%
\bibitem [{\citenamefont {Altarelli}\ and\ \citenamefont
  {Lipari}(1977)}]{AL77}%
  \BibitemOpen
  \bibfield  {author} {\bibinfo {author} {\bibfnamefont {M.}~\bibnamefont
  {Altarelli}}\ and\ \bibinfo {author} {\bibfnamefont {N.~O.}\ \bibnamefont
  {Lipari}},\ }\href {\doibase 10.1103/PhysRevB.15.4898} {\bibfield  {journal}
  {\bibinfo  {journal} {Phys. Rev. B}\ }\textbf {\bibinfo {volume} {15}},\
  \bibinfo {pages} {4898} (\bibinfo {year} {1977})}\BibitemShut {NoStop}%
\bibitem [{\citenamefont {French}\ \emph {et~al.}(2009)\citenamefont {French},
  \citenamefont {Schwartz}, \citenamefont {Stolz},\ and\ \citenamefont
  {Redmer}}]{FSSR09}%
  \BibitemOpen
  \bibfield  {author} {\bibinfo {author} {\bibfnamefont {M.}~\bibnamefont
  {French}}, \bibinfo {author} {\bibfnamefont {R.}~\bibnamefont {Schwartz}},
  \bibinfo {author} {\bibfnamefont {H.}~\bibnamefont {Stolz}}, \ and\ \bibinfo
  {author} {\bibfnamefont {R.}~\bibnamefont {Redmer}},\ }\href@noop {}
  {\bibfield  {journal} {\bibinfo  {journal} {J. Phys.: Condens. Matter}\
  }\textbf {\bibinfo {volume} {21}},\ \bibinfo {pages} {015502} (\bibinfo
  {year} {2009})}\BibitemShut {NoStop}%
\bibitem [{\citenamefont {Koster}\ \emph {et~al.}(1963)\citenamefont {Koster},
  \citenamefont {Dimmock}, \citenamefont {Wheeler},\ and\ \citenamefont
  {Statz}}]{KDWS63}%
  \BibitemOpen
  \bibfield  {author} {\bibinfo {author} {\bibfnamefont {G.}~\bibnamefont
  {Koster}}, \bibinfo {author} {\bibfnamefont {J.}~\bibnamefont {Dimmock}},
  \bibinfo {author} {\bibfnamefont {R.}~\bibnamefont {Wheeler}}, \ and\
  \bibinfo {author} {\bibfnamefont {H.}~\bibnamefont {Statz}},\ }\href@noop {}
  {\emph {\bibinfo {title} {Properties of the Thirty-Two Point Groups}}}\
  (\bibinfo  {publisher} {M.I.T. Press},\ \bibinfo {year} {1963})\BibitemShut
  {NoStop}%
\bibitem [{\citenamefont {Elliott}(1961)}]{Ell61}%
  \BibitemOpen
  \bibfield  {author} {\bibinfo {author} {\bibfnamefont {R.~J.}\ \bibnamefont
  {Elliott}},\ }\href {\doibase 10.1103/PhysRev.124.340} {\bibfield  {journal}
  {\bibinfo  {journal} {Phys. Rev.}\ }\textbf {\bibinfo {volume} {124}},\
  \bibinfo {pages} {340} (\bibinfo {year} {1961})}\BibitemShut {NoStop}%
\bibitem [{\citenamefont {Dahl}\ and\ \citenamefont {Switendick}(1966)}]{DS66}%
  \BibitemOpen
  \bibfield  {author} {\bibinfo {author} {\bibfnamefont {J.}~\bibnamefont
  {Dahl}}\ and\ \bibinfo {author} {\bibfnamefont {A.}~\bibnamefont
  {Switendick}},\ }\href {\doibase
  http://dx.doi.org/10.1016/0022-3697(66)90064-3} {\bibfield  {journal}
  {\bibinfo  {journal} {J. Phys. Chem. Solids}\ }\textbf {\bibinfo {volume}
  {27}},\ \bibinfo {pages} {931 } (\bibinfo {year} {1966})}\BibitemShut
  {NoStop}%
\bibitem [{\citenamefont {Pollmann}\ and\ \citenamefont
  {B\"uttner}(1977)}]{PB77}%
  \BibitemOpen
  \bibfield  {author} {\bibinfo {author} {\bibfnamefont {J.}~\bibnamefont
  {Pollmann}}\ and\ \bibinfo {author} {\bibfnamefont {H.}~\bibnamefont
  {B\"uttner}},\ }\href {\doibase 10.1103/PhysRevB.16.4480} {\bibfield
  {journal} {\bibinfo  {journal} {Phys. Rev. B}\ }\textbf {\bibinfo {volume}
  {16}},\ \bibinfo {pages} {4480} (\bibinfo {year} {1977})}\BibitemShut
  {NoStop}%
\bibitem [{\citenamefont {Saad}(1992)}]{Sa92}%
  \BibitemOpen
  \bibfield  {author} {\bibinfo {author} {\bibfnamefont {Y.}~\bibnamefont
  {Saad}},\ }\href {http://www-users.cs.umn.edu/$\sim$saad/books.html} {\emph
  {\bibinfo {title} {Numerical Methods for Large Eigenvalue Problems}}}\
  (\bibinfo  {publisher} {University Press},\ \bibinfo {address} {Manchester},\
  \bibinfo {year} {1992})\BibitemShut {NoStop}%
\bibitem [{\citenamefont {Sorensen}(2002)}]{So02}%
  \BibitemOpen
  \bibfield  {author} {\bibinfo {author} {\bibfnamefont {D.~C.}\ \bibnamefont
  {Sorensen}},\ }\href@noop {} {\bibfield  {journal} {\bibinfo  {journal} {Acta
  Numerica}\ }\textbf {\bibinfo {volume} {11}},\ \bibinfo {pages} {519}
  (\bibinfo {year} {2002})}\BibitemShut {NoStop}%
\bibitem [{\citenamefont {Alvermann}\ \emph {et~al.}(2011)\citenamefont
  {Alvermann}, \citenamefont {Littlewood},\ and\ \citenamefont
  {Fehske}}]{ALF11}%
  \BibitemOpen
  \bibfield  {author} {\bibinfo {author} {\bibfnamefont {A.}~\bibnamefont
  {Alvermann}}, \bibinfo {author} {\bibfnamefont {P.~B.}\ \bibnamefont
  {Littlewood}}, \ and\ \bibinfo {author} {\bibfnamefont {H.}~\bibnamefont
  {Fehske}},\ }\href {\doibase 10.1103/PhysRevB.84.035126} {\bibfield
  {journal} {\bibinfo  {journal} {Phys. Rev. B}\ }\textbf {\bibinfo {volume}
  {84}},\ \bibinfo {pages} {035126} (\bibinfo {year} {2011})}\BibitemShut
  {NoStop}%
\bibitem [{\citenamefont {Cho}(1976)}]{Cho76}%
  \BibitemOpen
  \bibfield  {author} {\bibinfo {author} {\bibfnamefont {K.}~\bibnamefont
  {Cho}},\ }\href {\doibase 10.1103/PhysRevB.14.4463} {\bibfield  {journal}
  {\bibinfo  {journal} {Phys. Rev. B}\ }\textbf {\bibinfo {volume} {14}},\
  \bibinfo {pages} {4463} (\bibinfo {year} {1976})}\BibitemShut {NoStop}%
\bibitem [{\citenamefont {Dasbach}\ \emph {et~al.}(2003)\citenamefont
  {Dasbach}, \citenamefont {Fr\"ohlich}, \citenamefont {Stolz}, \citenamefont
  {Klieber}, \citenamefont {Suter},\ and\ \citenamefont {Bayer}}]{DFSKSB03}%
  \BibitemOpen
  \bibfield  {author} {\bibinfo {author} {\bibfnamefont {G.}~\bibnamefont
  {Dasbach}}, \bibinfo {author} {\bibfnamefont {D.}~\bibnamefont {Fr\"ohlich}},
  \bibinfo {author} {\bibfnamefont {H.}~\bibnamefont {Stolz}}, \bibinfo
  {author} {\bibfnamefont {R.}~\bibnamefont {Klieber}}, \bibinfo {author}
  {\bibfnamefont {D.}~\bibnamefont {Suter}}, \ and\ \bibinfo {author}
  {\bibfnamefont {M.}~\bibnamefont {Bayer}},\ }\href {\doibase
  10.1103/PhysRevLett.91.107401} {\bibfield  {journal} {\bibinfo  {journal}
  {Phys. Rev. Lett.}\ }\textbf {\bibinfo {volume} {91}},\ \bibinfo {pages}
  {107401} (\bibinfo {year} {2003})}\BibitemShut {NoStop}%
\bibitem [{\citenamefont {Dasbach}\ \emph {et~al.}(2004)\citenamefont
  {Dasbach}, \citenamefont {Fr\"ohlich}, \citenamefont {Klieber}, \citenamefont
  {Suter}, \citenamefont {Bayer},\ and\ \citenamefont {Stolz}}]{DFKSBS04}%
  \BibitemOpen
  \bibfield  {author} {\bibinfo {author} {\bibfnamefont {G.}~\bibnamefont
  {Dasbach}}, \bibinfo {author} {\bibfnamefont {D.}~\bibnamefont {Fr\"ohlich}},
  \bibinfo {author} {\bibfnamefont {R.}~\bibnamefont {Klieber}}, \bibinfo
  {author} {\bibfnamefont {D.}~\bibnamefont {Suter}}, \bibinfo {author}
  {\bibfnamefont {M.}~\bibnamefont {Bayer}}, \ and\ \bibinfo {author}
  {\bibfnamefont {H.}~\bibnamefont {Stolz}},\ }\href {\doibase
  10.1103/PhysRevB.70.045206} {\bibfield  {journal} {\bibinfo  {journal} {Phys.
  Rev. B}\ }\textbf {\bibinfo {volume} {70}},\ \bibinfo {pages} {045206}
  (\bibinfo {year} {2004})}\BibitemShut {NoStop}%
\bibitem [{\citenamefont {Dasbach}\ \emph {et~al.}(2005)\citenamefont
  {Dasbach}, \citenamefont {Fr{\"o}hlich}, \citenamefont {Stolz}, \citenamefont
  {Klieber}, \citenamefont {Suter},\ and\ \citenamefont {Bayer}}]{DFSKSB05}%
  \BibitemOpen
  \bibfield  {author} {\bibinfo {author} {\bibfnamefont {G.}~\bibnamefont
  {Dasbach}}, \bibinfo {author} {\bibfnamefont {D.}~\bibnamefont
  {Fr{\"o}hlich}}, \bibinfo {author} {\bibfnamefont {H.}~\bibnamefont {Stolz}},
  \bibinfo {author} {\bibfnamefont {R.}~\bibnamefont {Klieber}}, \bibinfo
  {author} {\bibfnamefont {D.}~\bibnamefont {Suter}}, \ and\ \bibinfo {author}
  {\bibfnamefont {M.}~\bibnamefont {Bayer}},\ }\href {\doibase
  10.1002/pssc.200460331} {\bibfield  {journal} {\bibinfo  {journal} {Phys.
  Status Solidi C}\ }\textbf {\bibinfo {volume} {2}},\ \bibinfo {pages} {886}
  (\bibinfo {year} {2005})}\BibitemShut {NoStop}%
\bibitem [{\citenamefont {Trebin}(1977)}]{Tre77}%
  \BibitemOpen
  \bibfield  {author} {\bibinfo {author} {\bibfnamefont {H.-R.}\ \bibnamefont
  {Trebin}},\ }\href@noop {} {\bibfield  {journal} {\bibinfo  {journal} {Phys.
  Status Solidi B}\ }\textbf {\bibinfo {volume} {81}},\ \bibinfo {pages} {527}
  (\bibinfo {year} {1977})}\BibitemShut {NoStop}%
\bibitem [{\citenamefont {Waters}\ \emph {et~al.}(1980)\citenamefont {Waters},
  \citenamefont {Pollak}, \citenamefont {Bruce},\ and\ \citenamefont
  {Cummins}}]{WPBC80}%
  \BibitemOpen
  \bibfield  {author} {\bibinfo {author} {\bibfnamefont {R.~G.}\ \bibnamefont
  {Waters}}, \bibinfo {author} {\bibfnamefont {F.~H.}\ \bibnamefont {Pollak}},
  \bibinfo {author} {\bibfnamefont {R.~H.}\ \bibnamefont {Bruce}}, \ and\
  \bibinfo {author} {\bibfnamefont {H.~Z.}\ \bibnamefont {Cummins}},\
  }\href@noop {} {\bibfield  {journal} {\bibinfo  {journal} {Phys. Rev. B}\
  }\textbf {\bibinfo {volume} {21}},\ \bibinfo {pages} {1665} (\bibinfo {year}
  {1980})}\BibitemShut {NoStop}%
\end{thebibliography}
\end{document}